\documentclass[aps,pra,reprint,nofootinbib]{revtex4-1}
\usepackage{algorithm}
\usepackage{algorithmic}
\usepackage{longtable}     
\usepackage{epsf}          
\usepackage{bm}            
\usepackage[pdftex]{graphicx}
\usepackage[cmex10]{amsmath}
\usepackage{algorithmic}
\usepackage{color}
\usepackage{float}

\newcommand{\e}{\varepsilon}

\newcommand{\sd}{Schr\"{o}dinger }

\begin{document}

\title{Robust Control of Quantum Dynamics under Input and Parameter Uncertainty}

\author{Andrew Koswara, Vaibhav Bhutoria}
\affiliation{School of Chemical Engineering, Purdue University, West Lafayette, IN 47907 USA}
\author{Raj Chakrabarti}
\thanks{ Corresponding author: R. Chakrabarti (email: raj@pmc-group.com). Part of this work was carried out while RC was Professor of Chemical Engineering at Purdue University.}
\affiliation{Division of Fundamental Research, Chakrabarti Advanced Technology LLC, Mt. Laurel, NJ 08054 USA}

\begin{abstract}
Despite significant progress in theoretical and laboratory quantum control, engineering quantum systems remains principally challenging due to manifestation of noise and uncertainties associated with the field and Hamiltonian parameters.
In this paper, we extend and generalize the asymptotic quantum control robustness analysis method -- which provides more accurate estimates of quantum control objective moments than standard leading order techniques --
to diverse quantum observables, gates and moments thereof, and also introduce the Pontryagin Maximum Principle for quantum robust control.
In addition, we present a Pareto optimization framework for achieving robust control via evolutionary open loop (model-based) and closed loop (model-free) approaches with the mechanisms of robustness and convergence described using asymptotic quantum control robustness analysis.
In the open loop approach, a multiobjective genetic algorithm is used to obtain Pareto solutions in terms of the expectation and variance of the transition probability under Hamiltonian parameter uncertainty. The set of numerically determined solutions can then be used as a starting population for model-free learning control in a feedback loop. The closed loop approach utilizes real-coded genetic algorithm with adaptive exploration and exploitation operators in order to preserve solution diversity and dynamically optimize the transition probability in the presence of field noise. Together, these methods provide a foundation for high fidelity adaptive feedback control of quantum systems wherein open loop control predictions are iteratively improved based on data from closed loop experiments.  
\end{abstract}
\maketitle

\section{Introduction}
Quantum control is the application of control theory and optimization techniques to a quantum system coupled with an external field \cite{leth1977, bloe1984, warr1993}. Quantum control problems have been solved by both high-bandwidth pulse shaping in conjunction with optimization algorithms (quantum optimal control \cite{brif2010}) and modulation of a small number of field modes without the use of optimization algorithms to induce population transfer into the desired state (e.g., stimulated Raman adiabatic passage or STIRAP \cite{kuklinski1989,gaubatz1990,shorebergmann1991,kobrakrice1998}).
Quantum control systems have been theoretically analyzed for \textit{controllability} \cite{levi2001,brif2010}, as well as for their \emph{control landscapes} \cite{rabi2006, roth2006, shen2006, chak2007}. The former studies and determines whether a manipulated field exists and is able to coherently direct the evolution of the system from an initial to an arbitrary final state after a sufficiently long time. On the other hand, the latter is the map between the manipulated field parameters and the control performance measure which has been analytically shown to harbor no local traps given a controllable Hamiltonian and unconstrained field parameters. The analysis provides insights and explanations for the successes of laboratory quantum learning control, especially in the manipulation of molecular evolution along its excited potential energy surface such as in photodissociation reaction \cite{brix2001}, optimal dynamic discrimination \cite{peter2010}, and in excitonic energy transfer \cite{scho2010}. Nevertheless, manipulating quantum dynamics is still largely difficult for both theoretician and experimentalist. One principal reason is because the knowledge of a quantum system's Hamiltonian becomes exponentially more complex as it increases in size and, as a result, the Hamiltonian parameters are not known exactly from ab initio calculations. In addition, due to the nonlinear dynamics of quantum systems, estimation of Hamiltonian parameters based on experimental measurements leads to an uncertainty distribution of parameter values \cite{youngwhaley2009,chak2012}. Disturbances also manifest in the control field parameters due to limited precision of field-shaping devices and fluctuations of the field source \cite{wein2000, jian2007}.  Hence, the primary challenge in \emph{robust} quantum control is to develop an experimentally implementable robust control strategy, which also agrees with the first-principles model. Recent work on robustness of controlled quantum dynamics showed that in order to maximize the robustness of a control field in terms of its resistance to noise and uncertainty in the control and system parameters, the effect of parameter distribution on destructive interference between different order transitions must be minimized \cite{akos2014}. Whereas model-based quantum control solutions are at present seldom successfully implemented in experiments, appropriate quantum control robustness analysis and optimization techniques in conjunction with accurate estimates of Hamiltonian parameter distributions and noise spectra will improve the fidelity of a wide variety of previously reported model-based control solutions. The control of more complex quantum systems will also benefit from integration of robust model-based and model-free control strategies.

Inspired by well-developed concepts in the field of classical control engineering, quantum control \textit{robustness} is described in terms of the effect of disturbances in the control field and the system's Hamiltonian parameters on the different moments of the quantum control objective \cite{akos2014,nagy2004,ferr1997}. The control objectives may be formulated mainly for two applications: the first is to reach a target dynamical propagator or \textit{quantum gate} for applications in quantum computation and information processing, and the second is to maximize a \textit{quantum observable}, such as in coherent control of chemical reactions or in photosynthetic energy transfer. The concept of robustness is essential in ensuring control objectives are satisfied in the presence of parameter distribution \cite{ma1999,fan1991,braa1996}. Historically, no prior work has studied the mechanism of robustness in high-fidelity quantum control. In addition, while evolutionary algorithms have been successfully implemented in laboratory quantum control in obtaining optimal control profiles \cite{brix2001,peter2010,scho2010}, analysis of the performance of evolutionary algorithms in terms of their ability to Pareto optimize moments of the control objective in the presence of field noise and Hamiltonian uncertainty is a relatively new topic of study. Moreover, experimental quantum learning control has primarily been applied to the problem of population transfer between pure states \cite{brif2010}, whereas applications to control of mixed states and quantum gates are more challenging to achieve without employing model-based robust control \cite{kosu2013}.

A number of robust quantum control theories and algorithms, which are rooted in their classical counterparts, have been considered for different control settings \cite{hornungmotzkus2002,barteltroth2005,hock2014,kosu2013,grac2012,mabu2005,rabi2018},
including robust control design via leading Taylor series \cite{hock2014,kosu2013}, such as in minimax or worst-case formulations \cite{nagy2004}. However, these methods are limited in their accuracy. 
 Recent work from the authors has also shown that robustness of controlled quantum dynamics may be understood in terms of the explicit contribution of control and system parameters to the dynamics, which can be decomposed into \emph{quantum pathways} \cite{akos2014}. In contrast to the leading-order Taylor expansion, this method is capable of exactly determining the robustness of a control field in terms of the robust combination and interferences between all significant pathways.  In this paper, we generalize this theory to arbitrary quantum observables, gates and moments thereof, describe methods for dimensionality reduction through the identification of significant quantum pathways, and also introduce the Pontryagin Maximum Principle for the optimality conditions for quantum control objective moments in the presence of noise and uncertainty. Based on this theory of quantum robust control optimization, evolutionary algorithm-based quantum robust control design is proposed. The proposed control strategies are classified into two different control paradigms, namely model-based and model-free robust control, which have the potential for integration. 
The objectives of the robust control design are two-fold; the first is to systematically explore robust quantum control solutions and the second is to overcome the gap between model-based and model-free quantum control in the laboratory. Following from recent advances in quantum optimal and robust control theory, the performance of the robust control design can be rigorously analyzed in terms of the quantum control landscape, quantum control pathways, and their interferences.

The paper is organized as follows: the generalized theory for quantum observable and gate control robustness analysis, and the Pontryagin Maximum Principle for control of moments of these objectives, are first presented in Section \ref{Quantum_Robust_Control_Theory}. 
Based on the foundation for robust optimization of moments of quantum observables, model-based and model-free quantum optimal and robust control via evolutionary approaches are discussed in Section \ref{Robust_Control_via_Evolutionary_Approach}. Here, the details of the algorithms' implementation are outlined with emphasis on how the quantum control theory and robustness analysis are used to aid the optimization settings. The implementation of  different types of genetic algorithms (GAs) and their utilities in different robust control scenarios are also described. In Section \ref{Results_and_Discussions}, the results of the different control implementations on a model quantum system is described in details in terms of quantum pathways and their interferences. The paper concludes with the summary of current work and how the results can be generalized and used in future work in Section \ref{Summary_and_Outlook}.

\section{Quantum Robust Control Theory}
\label{Quantum_Robust_Control_Theory}
\subsection{Quantum Optimal Control Landscape}
The dynamics of a quantum system coupled with an external electric field is described by the \sd equation:
\begin{gather}
	\label{SDunit}
	\frac{dU(t)}{dt}=-\frac{i}{\hbar}\left(H_0-\mu\varepsilon(t)\right)U(t),~U(0)=I,
\end{gather}
where $H_0$ is the time-independent Hamiltonian of the system, $\mu$ the dipole moment, $\varepsilon(t)$ the time dependent field, and $U(t)$ denotes the dynamical or \emph{unitary} propagator, i.e. $|\psi(t)\rangle=U(t)\psi(0)\rangle$. Here, the quantum state $|\psi(t)\rangle$ is said to be in a Hilbert space of finite dimension $N$ and $\mu$ is a real symmetric matrix of the same dimension. The quantum states of interest depend on the analytical form of $H_0$ and include vibrational, rotational and spin states common in quantum control examples. In the context of optimal control, the state of system is $U(t),\forall~t \in [0,T]$ and is the \textit{control variable}. The quantum system is controlled by a time-dependent field $\varepsilon(t)$, which is the \textit{manipulated variable}. In order to simplify the ensuing analysis, the notation for the interaction Hamiltonian $H_I(t) = e^{\frac{i}{\hbar}H_0t}\{-\mu\e(t)\}e^{-\frac{i}{\hbar}H_0t}$ is used, giving:
\begin{gather}
	\frac{dU_{I}(t)}{dt}=-\frac{i}{\hbar}H_{I}(t)U_{I}(t),~ U_I(t)=e^{\frac{i}{\hbar}H_0t}U(t).
	\label{SDunitinter}
\end{gather}	
The subscript $I$ is subsequently dropped from the description of the unitary propagator in the interaction picture for convenience. The objective of the optimization is to achieve desired control objective obeying the \textit{dynamical constraint} described by (\ref{SDunitinter}) as well as limitations in the laboratory. This can be succinctly described in the following control formulation:
\begin{gather}
	\underset{\varepsilon(\cdot)}{\mathrm{min}}~F(U(T)), \nonumber \\
	\hspace{-2.5in}sbj~to: \nonumber \\
	\nonumber \\
	\frac{dU(t)}{dt}=-\frac{i}{\hbar}H(t)U(t),~ U(0)=I, \nonumber \\
	\varepsilon(t)=\sum_k^K A_k~cos(\omega_kt + \phi_k), \nonumber \\
	A_{min} \leq A_k \leq A_{max}, \nonumber \\
	0 \leq \phi_k \leq 2\pi, \nonumber \\
	\omega_{min} \leq \omega_k \leq \omega_{max}, \nonumber \\
	0 \leq t \leq T.
	\label{landscape}
\end{gather}
where $F(U(T))$ is the \textit{cost function} and depends on the quantum control objective and the final state of the system, and the control field is expressed as combination of linearly polarized cosine waves. In this work, two performance measures are relevant. The first is related to optimization of a quantum expectation of an observable $\Theta$ given an initial density matrix $\rho_0$, namely $\operatorname{Tr}\left[U(T) \rho_{0} U^{\dagger}(T) \Theta\right]$, a special case of which is the transition probability between an initial pure state $|i\rangle$ and a final pure state $|j\rangle$:

\begin{gather}
F(U(T))=|U_{ji}(T)|^2-1,  \\
0 \leq |U_{ji}(T)|^2 \leq 1.\nonumber
\label{cost}
\end{gather}
and the second is related to optimization of the fidelity of a quantum gate in quantum information processing \cite{kosu2013}:
\begin{gather}
F(U(T))=|U(T)-W|^2
\label{cost}
\end{gather}
where $W$ is the target quantum gate. The notation $J(\varepsilon(t))$ is used to denote a generic quantum control objective as a functional on the space of input control fields.

%
\begin{figure*}
	\centering
	\includegraphics[width=18cm]{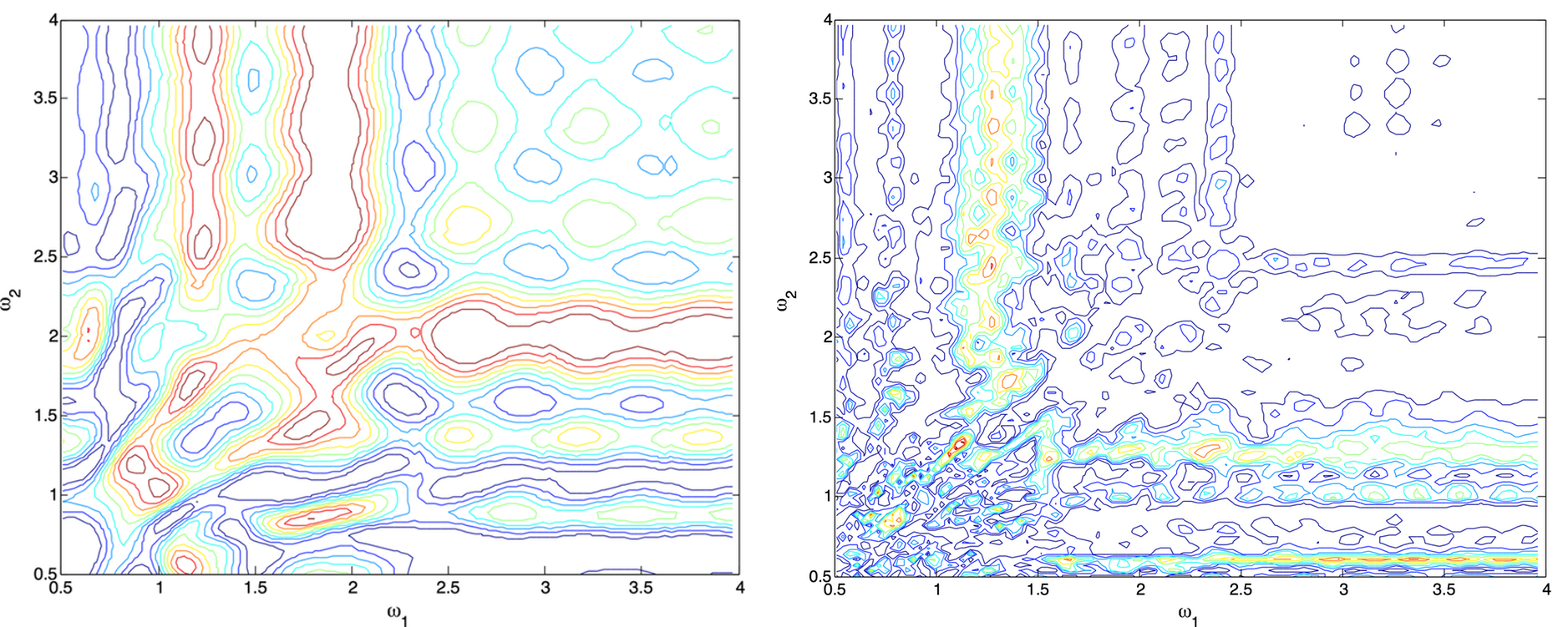}
	\caption{\label{qcland} Quantum control landscape as a function of two frequency modes. (left) Nominal landscape of an artificial four-level system with bounded field duration and amplitude. (right) Landscape of a five-level system with reduced field duration and amplitude. The contour plots show how the landscape changes in terms of the shift in optimal points and reduced multimodality.}
\end{figure*}
%

The quantum control landscape \cite{chak2007} determines how the manipulated parameters map to the performance measure.
This landscape has been analyzed in terms of its topology, geometry and search complexities given unconstrained field parameters \cite{chak2007}. The analysis demonstrates that all critical points are either global or saddle and that there exists a multiplicity of control solutions. In contrast, however, optimal control is not always achieved in the laboratory due to practical constraints and lack of robustness of the real quantum system. For instance, bounds on the field's spectral parameters or field duration and the presence of uncertainties in the system can significantly affect the control landscape.

As mentioned previously, for control with infinitely flexible constraints, the landscape is known to be multimodal and trap-free \cite{chak2007}. However, as constraints such as reduction in the field duration and the number of frequency modes which can be manipulated to control the system are imposed on the optimization, the constrained landscape reveals local traps (Figure \ref{qcland} (Left)), which depicts a landscape in 2D space with respect to the first two field frequency modes $\omega_1$ and $\omega_2$). %
Moreover, Figure \ref{qcland}  (Right)  
shows how the number of sub-optimal points significantly increases and the level-set reduces as more limitations are imposed on the system. Additionally, when the system parameters are varied due to uncertainties $\delta$, the shape of the landscape can change dramatically, shifting and decreasing the location and the number of optimal points. 
Note that given $\delta$ $\sim$ $N(0,\sigma^{2})$  $\mathrm{E}[P_{ji}]$ would not be equal to $P_{ji}$ at the nominal values of the parameters unless it is linear. Hence, robust solutions are generally not an optimum of the original landscape, and there may be a trade-off between performance quality and robustness of optimal solutions. Additional local optima may be introduced in the fitness landscapes which may be suboptimal but more robust. Thus, a robust optimization problem can become a multimodal optimization problem, in which it is essential to include $\delta$ in the optimization and modify the robust control strategies such that exploration of both global and local optima are maximized.
\subsection{Asymptotic Robustness of Quantum Control}
\label{Robustness_of_Quantum_Control}
The asymptotic robustness 
of quantum control \cite{akos2014} of transition amplitude and probability can be described in terms of the different moments of these quantities
broken down into \emph{quantum pathways}. These concepts can also be generalized to arbitrary quantum observables and gates.

The controlled transition amplitude $U_{ji}(T)$ is calculated as an infinite sum of Dyson terms, i.e. $U_{ji}(T)=\sum_{m}^\infty U_{ji}^m(T)$ 
\cite{mitr2003}, where the $m$-th order term is defined as:
\begin{align}	
	& U_{ji}^m(T)= \langle j|\left(-\frac{\imath}{\hbar}\right)^m\int_0^T\int_0^{t_2} \cdots \int_0^{t_{m-1}}  H_I(t_{1})~H_I(t_{2}) \cdots \nonumber \\
	&~~~~~H_I(t_m) ~dt_1~dt_2\cdots dt_{m}| i \rangle.
	\label{DysSer}
\end{align}
Each of the $m$-th order term can in turn be expanded in terms of the field's spectral parameters:
\begin{align}
	U_{ji}^m&(T) = \left(\frac{\imath}{\hbar}\right)^m \mu_{jl_{m-1}} \cdots \mu_{l_1i} \times \nonumber \\
	&\sum_{k_{m}=1}^K A_{k_m} \int_0^T e^{\imath\omega_{jl_{m-1}}t_{m}} \cos(\omega_{k_m}t_{m} + \phi(\omega_{k_m})) \times \cdots \nonumber \\
	&\times \sum_{k_{1}=1}^K A_{k_1} \int_0^{t_{2}}e^{\imath\omega_{l_1i}t_{1}} \cos(\omega_{k_1}t_{1} + \phi(\omega_{k_1}))~ dt_1\cdots dt_{m}.
	\label{dysonserparam}
\end{align}
where $1 \leq l_i, l_{m-1} \leq N$ indicate the transitions allowed by the dipole moment. If we use the notation $U_{ji}^m(T)$ to denote the $m$-th order term in the series above, the total transition amplitude between an initial state $|i\rangle$ and a final state $|j\rangle$ at time $T$ can be expressed as:
\begin{align}
	U_{ji}(T)=\sum_m^\infty U_{ji}^m(T),
	\label{transamp}
\end{align}
and the transition probability as:
\begin{align}
	P_{ji}&=
	\sum_{m}^\infty \left|U_{ji}^m(T)\right|^2+2\sum_{m'<m}\mathrm{Re}\left\{(U_{ji}^{m}(T))(U_{ji}^{m'}(T))^*\right\}.
	\label{transprob}
\end{align}
The above expression demonstrates the property of quantum interferences due to the presence of coherence terms $(U_{ji}^mU_{ji}^{m'*})$.  \emph{Constructive} interference takes place when the sum of coherences is larger than 0, and \emph{destructive} interference when the sum is less than 0.

Following from recent work in quantum control robustness analysis \cite{akos2014}, each order transition amplitude may be described as a sum of quantum pathways:
\begin{align}
	U_{ji}^m(T) = \sum_{\vec{\alpha} \in \mathcal{M}} U_{ji}\left(T,\vec{\alpha}\right),
	\label{transamppath}
\end{align}
where the notation $\vec \alpha \in \mathcal{M}$ represents pathways belonging to a particular order $m$. In the presence of input noise or uncertainty in system parameters, a \emph{normalized} quantum pathway can be further defined:
\begin{gather}
	c_{\vec{\alpha}} = \frac{U_{ji}(T,\vec{\alpha})}{\prod_{\vec \alpha} \theta_k^{\alpha_k}}.
	\label{amppathnorm}
\end{gather}
where $\theta$ denotes a vector of noisy or uncertain parameters (such as field mode amplitudes, frequencies or phases for input noise, or parameters of the control Hamiltonian matrix $\mu$ for Hamiltonian uncertainty).
In turn, assuming for simplicity that the parameters $\theta_k$  are independent random variables, the expected transition amplitude is:
\begin{align}
	\mathrm{E}[U_{ji}(T)]&=\mathrm{E}\left[\sum_mU_{ji}^m(T)\right] \\ \nonumber
		&=\sum_{m}\sum_{\vec{\alpha}\in \mathcal{M}}c_{\vec{\alpha}}\prod_{k}^K\mathrm{E}[\theta_k^{\alpha_k}],
\end{align}
and, the variance:
\begin{align}
	\mathrm{var}&\left(\mathrm{Re, Im}\left\{U_{ji}(T)\right\}\right)= \nonumber \\
	&\hspace{0.1in}\sum_{\vec{\alpha}}\left(\mathrm{Re,Im}\left\{c_{\vec{\alpha}}\right\}\right)^2 \left(\prod_{k}^K\mathrm{E}[\theta_k^{2\alpha_k}]- \prod_{k}^K\mathrm{E}^2[\theta_k^{\alpha_k}]\right)+ \nonumber \\
	&~~~~~~2\sum_{\vec{\alpha}'< \vec{\alpha}}\mathrm{Re,Im}\left\{c_{\vec{\alpha}}\right\} \mathrm{Re,Im}\left\{c_{\vec{\alpha}'}\right\} \times \nonumber \\
	&~~~~~~\left(\prod_{k}^K\mathrm{E}[\theta_k^{\alpha_k'+\alpha_k}]-\prod_{k}^K\mathrm{E}[\theta_k^{\alpha_k'}]\prod_{k}^K\mathrm{E}[\theta_k^{\alpha_k}] \right).
	\label{vartransprob}
\end{align}

The expected transition probability and its variance are similarly derived:
\begin{align}
	\mathrm{E}&\left[P_{ji}(T)\right]=\mathrm{E}\left[\sum_m\left|U_{ji}^m\right|^2\right]+ \nonumber \\
	&\mathrm{E}\left[2\sum_{m'< m}\mathrm{Re}\left\{(U_{ji}^{m'})(U_{ji}^{m})^*\right\}\right].
	\label{momtransprob}		
\end{align}
Here, $\theta_k$ is the noisy or uncertain parameters (i.e. amplitude, phase or dipole moments), and therefore, also depends on the types of pathways being analyzed. Using this notation, the expected transition probability may be expressed as:
\begin{align}
	\mathrm{E}&\left[P_{ji}(T)\right] = \mathrm{E}\left[\left|\sum_{\vec{\alpha}} c_{\vec{\alpha}}\prod_{k}^K \theta_k^{\alpha_k}\right|^2\right] \nonumber \\
	&=\hspace{0.1in}\sum_{\vec{\alpha}} \left|c_{\vec{\alpha}}\right|^2 \prod_{k}^K \mathrm{E}[\theta_k^{2\alpha_k}] + \nonumber \\
	&2\sum_{\vec{\alpha}' < \vec{\alpha}} \mathrm{Re}\left\{c_{\vec{\alpha}'} c^*_{\vec{\alpha}}~\prod_{k}^K\mathrm{E}[\theta_k^{\alpha_k'+\alpha_k}]\right\},
\end{align}

In order to compute the $\mathrm{E}[P_{ji}]$, the individual pathway terms $U_{ji}(T,\vec \alpha)$ are first calculated. This is accomplished using method of \emph{pathway encoding/decoding} \cite{mitr2003, mitr2008}, of which Fourier encoding is the most common type. Additionally, the higher moments of noisy or uncertain manipulated input and system parameters must first be computed. These terms can be provided in closed form for any uncertainty distribution for which there exists an analytical Fourier transform:
\begin{align}
	\phi(s) &= \langle e^{\imath \theta s} \rangle = \int_{-\infty}^{\infty} p(\theta) e^{\imath \theta s}~d\theta,
\end{align}
Here, $\phi(s)$ (not a control field phase) is referred to as the \emph{characteristic moment-generating} function of the probability distribution of the manipulated input or system parameter $p(\theta)$. If $\phi(s)$ is available in closed form, the moments of $\theta$ can be obtained via:
\begin{align}
	\langle \theta^k \rangle  &= (-\imath)^k \left[\frac{\partial^k}{\partial s^k}\phi(s)\right]_{s=0}
	\label{moment}
\end{align}
The asymptotic robustness analysis method is applicable to fully general field noise and Hamiltonian parameter uncertainty distributions, including any form of statsitical interdependence,
whereas leading order methods can only use the first and second moments of noise/uncertainty in the calculation of quantum control objective moments.
Moreover, leading order methods can only provide $E[J]$ to second order and var$(J)$ to first order \cite{nagy2004}.
Numerically, the moments derived above can be compared to the leading order Taylor expansion approximations for the first and second moments of state-to-state population transfer reported in \cite{akos2014}.
In general, there is no a priori reason to believe the leading order moment calculations are sufficiently accurate for model-based control applications.
In particular, assuming Gaussian noise/parameter uncertainty, $\delta  E[J]=0$ to first order \cite{akos2014}. 

\subsection{Robustness of Significant Quantum Pathways}
\label{subsetencoding}
In previous work \cite{akos2014}, Fourier encoding of pathways was performed for a full set of uncertain or noisy parameters. We refer the reader to the reference for details of the approach. However, there are cases where only a subset of parameters are relevant, for example those involving uncertain dipole moments or noisy field modes that have a significant impact on controlled dynamics. Subset encoding is especially useful when computational complexity is an issue and only mechanistic and robustness analysis of the most significant subset of parameters are desired. The method of subset encoding and decoding for a subset of pathways is described in this section. Given that $\theta_k$ specifies the uncertain system or input parameter, the encoding of subset of parameters subject to uncertainty is as follows:
$$\theta_k \rightarrow \theta_k\exp(i\gamma_k s), k = 1,...,n;~ n \leq n_{\max}.$$
where $n$ denotes the number of uncertain parameters and $n_{\max}$ the total number of parameters. Without loss of generality, the uncertain parameters are indexed such that they are the first $n$ parameters. In this way, all terms containing $\theta_1^{\alpha_1}....\theta_n^{\alpha_n}$, can be extracted via encoded transition amplitude $U_{ji}(T,\gamma=\alpha_1\gamma_1 + ... + \alpha_n \gamma_n)$. 
\begin{align*}
U_{ji}&(T,\gamma=\alpha_1\gamma_{1} + ... + \alpha_n \gamma_{n}) = \\
	&=\theta_{1}^{\alpha_1}...\theta_{n}^{\alpha_n}\biggr[(i^{m1})\sum_{k_{1}=n+1}^{n_{\max}} \theta_{k_1}...\sum_{k_{m1}=n+1}^{n_{\max}} \theta_{k_{m1}}...+\\
	&~~~~~~~+(i^{m_{2}})\sum_{k_{1}=n+1}^{n_{\max}}\theta_{k_{1}}...\sum_{k_{m2}=n+1}^{n_{\max}}\theta_{k_{m_{2}}}...+...\biggr]
\end{align*}
It is important to note, however, that in the full-encoding method there was no sum over unencoded parameters. In addition, the coefficient $c_{\alpha_1,...,\alpha_n}$, which is written without specification of a constraint on the $\alpha_i$'s, contains contributions from many different orders. Hence, for subset encoding $m = \sum_i \alpha_i$ is not a Dyson series order. It, however, plays an important role in determining the effect of noise on the pathway norm and interferences involving that pathway, since the ratio $\frac{\mathrm{E}\left[\prod_k \theta_k\right]}{\prod_k \theta_k}$ depends on $m$.

Significant quantum pathways can be identified with the use of  leading order sensitivity analysis methods. Consider, for example, the problem of minimizing var$(J)$ in the presence of field noise when 
$J_{\text{nom}}$ has been maximized  (the ``top" of the quantum control landscape  \cite{rabi2006, roth2006, shen2006, chak2007}), where $J_{\text{nom}}$ denotes the nominal value of $J$ for a given control field and estimated value of the time-independent Hamiltonian parameters in the absence of uncertainty/noise.
The most common leading order approximation for robust control is the first-order approximation for which there is an analytical distribution, which is Gaussian \cite{nagy2004}. Here
$\mathrm{E}[\delta J]$, where $\delta J \equiv J – J_{\text{nom}}$, is 0 for Gaussian noise/uncertainty, which results in the distribution for $J$ being centered at  $J_{\text{nom}}$ \cite{akos2014}.  To first-order, $\mathrm{var}~J \approx \int_0^{\Omega} \int_0^{\Omega} \frac{\delta J}{\delta \varepsilon(\omega')}\mathrm{Acf(\omega',\omega)} \frac{\delta J}{\delta \varepsilon(\omega)}~d\omega' d\omega$, where $\frac{\delta J}{\delta \varepsilon(\omega)}$  denotes the frequency domain gradient vector and $\mathrm{Acf(\omega',\omega)}$ denotes the frequency domain autocovariance function.
However, at the top of the landscape, this variance approximation
is also 0 
because $\frac{\delta J}{\delta \varepsilon(\omega)}$  is identically 0 there. 
Thus  at the extrema of the nominal $J_{\text{nom}}$ landscape the $J_{\text{nom}},~\mathrm{var}(J)$ and $J_{\text{nom}},~\mathrm{E}[J]$ Pareto fronts cannot be sampled using the standard first-order Taylor approximation to the distribution of $J$.

As such, at the extrema of the  $J_{\text{nom}}$ landscape, a second-order Taylor approximation must be used for both $\mathrm{E}[\delta J]$ \cite{hock2014} and $ \mathrm{var}(J)$. These are both expressed
in terms of the Hessian kernel of $J$ with respect to the uncertain/noisy parameters.
Considering the leading (second-) order approximation for $\mathrm{E}[\delta J]$ at such extrema of the quantum control landscape, 
we have
\begin{align}\label{autocov}
\mathrm{E}[\delta J]  &\approx\frac{1}{2}  \int_0^\Omega\int_0^\Omega\mathrm{Acf}(\omega’,\omega) \mathrm{Hess}(\omega’,\omega)  d\omega’ d\omega, 
\end{align}
where $\mathrm{Hess}$ denotes the Hessian kernel of the quantum control objective functional $J(\varepsilon)$.
Hess is a finite-rank kernel of rank $2N-2$ for the problem of state to state population transfer on Hilbert space dimension $N$ \cite{brif2010}.
At the optima of the nominal quantum control landscape, for var$(J)$ (or $\mathrm{E}[\delta J]$) calculation, only those uncertain/noisy parameters corresponding to terms in the integral (\ref{autocov}) above a specified magnitude (of which there are at most $2N-2$ in the basis where the Hessian is diagonal) can be included in the asymptotic moment calculation.  
Thus at the ``top"  of the quantum control landscape, employing a Fourier basis wherein $\mathrm{Hess}(\omega',\omega)$ is diagonal
, along with expression of the noise/uncertainty distribution on these transformed parameters, enables a substantial reduction in computational expense of the asymptotic var$(J)$ (or E$[\delta J]$) calculation, through the use of at most $2N-2$ encoding frequencies for population transfer problems (similarly for Hamiltonian parameter uncertainty).

We denote by $\Gamma=\left\{0, \ldots, \gamma_{f}\right\}$ the set of encoding frequencies corresponding to the significant pathways, and where the upper limit of integration is set to $s_f$ for a specified error tolerance on the moment calculation. Additional methods for dimensionality reduction of asymptotic quantum control robustness calculations exist, and will be presented in future work.

\subsection{Representation of Quantum Interference Moments}
\label{Polar_Representation_of_Quantum_Interferences}
Adding onto the description of quantum pathways in the complex plane, the description of quantum pathway interference moments can be further elaborated in the polar representation. Using the notation $A_{\vec \alpha} \equiv |U_{ji}^{\vec \alpha}|$ to represent a pathway and $\phi_{\vec \alpha} = \tan^{-1}\frac{\mathrm{Im}(c_{\vec \alpha})}{\mathrm{Re}(c_{\vec \alpha})}$ as the phase angle between two pathways, the transition probability may be rewritten as:
\begin{align*}
P_{ji}&=\sum_{\vec \alpha} |U_{ji}^{\vec \alpha}|^2 + 2\mathrm{Re}\sum_{\vec \alpha' < \vec \alpha'}U_{ji}^{\vec \alpha}U_{ji}^{\vec \alpha',*} \\
	&= \sum_{\vec \alpha} A_{\vec \alpha}^2 +   2\sum_{\vec \alpha' < \vec \alpha'} A_{\vec \alpha} A_{\vec \alpha'}\cos\left(\phi_{\vec \alpha} - \phi_{\vec       \alpha'}\right)                                                                                                                                                  \end{align*}
By definition, constructive interference corresponds to $\cos\left(\phi_{\vec \alpha} - \phi_{\vec \alpha'}\right) > 0$ and destructive interference to $\cos\left(\phi_{\vec \alpha} - \phi_{\vec \alpha'}\right) < 0$. 
%
%
In order to depict the contribution of interferences between different pathways to the transition probability, the norm of the sum of two pathways $U_{ji}^{\vec \alpha}$ and $U_{ji}^{\vec \alpha'}$ is considered:
\begin{align*}
&|U_{ji}^{\vec \alpha}+  U_{ji}^{\vec \alpha'}|^2 = |U_{ji}^{\vec \alpha}|^2 +  |U_{ji}^{\vec \alpha'}|^2 + 2\left(U_{ji}^{\vec \alpha} \cdot U_{ji}^{\vec \alpha'} \right)\\                                                                                                                                                        	
&~~=  |U_{ji}^{\vec \alpha}|^2 +  |U_{ji}^{\vec \alpha'}|^2 \\
&~~~~~+ 2 \left(\mathrm{Re}\left\{U_{ji}^{\vec \alpha}\right\}\mathrm{Re}\left\{U_{ji}^{\vec \alpha'}\right\} + \mathrm{Im}\left\{U_{ji}^{\vec \alpha}\right\}\mathrm{Im}\left\{U_{ji}^{\vec \alpha'}\right\}\right)
\end{align*}
such that the interference term is naturally interpreted in terms of the dot product of the pathway vectors. If we now define $A_{\vec \alpha} \equiv |c_{\vec \alpha}|$ as the normalized quantum pathway and take into account noise/uncertainty associated with the control and input parameters, the expected transition probability is described as:
\begin{align*}
\mathrm{E}[P_{ji}]&=\sum_{\vec \alpha} \mathrm{E}\left[|U_{ji}^{\vec \alpha}|^2\right] + \sum_{\vec \alpha' < \vec \alpha'}\mathrm{E}\left[2\mathrm{Re}~U_{ji}^{\vec                 \alpha}U_{ji}^{\vec \alpha',*}\right] = \\                                                                                                                     
&= \sum_{\vec \alpha} A_{\vec \alpha}^2\mathrm{E}\left[\prod_k \theta_k^{2\alpha_k}\right]  +   \\
&~~~~~~~+2\sum_{\vec \alpha' < \vec \alpha'} A_{\vec \alpha} A_{\vec     \alpha'} \mathrm{Re}~e^{\left(\imath(\phi_{\vec\alpha} - \phi_{\vec \alpha'})\right)}\mathrm{E}\left[\prod_k \theta_k^{\alpha_k+\alpha_k'}\right]\\
&= \sum_{\vec \alpha} A_{\vec \alpha}^2\mathrm{E}\left[\prod_k \theta_k^{2\alpha_k}\right]  +   \\
&~~~~~~~+2\sum_{\vec \alpha' < \vec \alpha'} A_{\vec \alpha} A_{\vec     \alpha'}\cos\left(\phi_{\vec \alpha} - \phi_{\vec \alpha'}\right) \mathrm{E}\left[\prod_k \theta_k^{\alpha_k+\alpha_k'}\right]
\end{align*}
Hence the dot product between the $c_{\alpha}$ vectors is retained, but note that the expectation of the product of parameters is not equal to the product of their expectations.

\subsection{Pontryagin Maximum Principle for Quantum Robust Control}

Whereas  \cite{akos2014} introduced methods for quantum control robustness analysis, we now extend this theory to identify optimality conditions for quantum robust control that establish a foundation for robust control optimization.
Regardless of the algorithm used for control optimization, the optimality conditions we derive below apply at the Pareto optima for robust control. In particular, conditions can be derived for both maximization of the expectation of a quantum control objective
and minimization of the variance of a quantum control objective. Together with deterministic quantum control optimality conditions \cite{moore2011}, these criteria define the properties of optima for any of the common quantum robust control Pareto fronts comprised
of combinations of $\mathrm{E}[J], \mathrm{var}~J$, and $J_{\text{nom}}$.

The costate equation, which imposes the dynamical constraints in control optimization, and the PMP (Pontryagin Maximum Principle; first-order conditions for optimality) for quantum robust control can be derived analogously to the deterministic quantum control PMP \cite{moore2011}.
However, the state and costate ($\Phi$) equations for the robust quantum control PMP are partial differential equations. Both state and costate are functions of $t$ and the timelike variable $s$.

A generalized expression for the moments of quantum observables or gates is required in order to derive the PMP. With $F(U(T))$ representing the quantum observable expectation
value or gate fidelity, we have for the first moment:
\[
\begin{aligned}
&\mathrm{E}\left[F(U(T))\right]=
\sum_{\gamma^{\prime} \in \Gamma} \int_{0}^{s_{f}} F(U(T, s)) \exp (-i \gamma s) d s \\ 
&~~~\cdot \delta\left(\gamma, \gamma^{\prime}\right) \cdot \frac{\mathrm{E}\left[\prod_{k} \theta_k^{\alpha_{k}\left(\gamma^{\prime}\right)}\right]}{\prod \theta_k^{\alpha_{k}\left(\gamma^{\prime}\right)}}
\end{aligned}
\].

Using the PMP-Hamiltonian function $\mathcal{H}=\left\langle\Phi(t, s),-\frac{i}{\hbar} H(t, s) U(t, s)\right\rangle$ we can obtain first-order conditions for optimality of moments. In deterministic quantum control, the Lagrangian $\bar J$ (objective function augmented by Lagrange multiplier constraints) is expressed in terms of $\mathcal{H}$  and $\left\langle\Phi(t), \frac{d U(t)}{d t}\right\rangle$ is integrated by parts to obtain  \cite{moore2011}
\[
\begin{aligned}
&\bar{J}=F(U(T))-\operatorname{Tr}(\Phi^{\dag}(T) U(T))+\operatorname{Tr}(\Phi^{\dag}(0) U(0))+\\
&~~\int_{0}^{T} \mathcal{H}(U(t), \Phi(t), \varepsilon(t))+\operatorname{Tr}\left(\frac{d \Phi^{\dag}(t)}{d t} U(t)\right) d t
\end{aligned}
\]

In robust quantum control, we are interested in $\mathrm{E}[\delta \bar{J}]$.
The costate equation for quantum robust control is the following partial differential equation in $t, s$:  
\[
\frac{\partial}{\partial t} \Phi(t, s)=-\frac{i}{\hbar} H(t, s) \Phi(t, s)
\]
subject to a terminal boundary condition on $\Phi(T, s)$ to be derived below.
The expression for $\mathrm{E}[\delta \bar{J}]$ can be evaluated by first writing the s-evolved Lagrangian $\bar{J}$ and then considering its first-order variation
\begin{widetext}
\[
\begin{aligned}
\delta \bar{J}(s)=\operatorname{Tr}\left(\left[\nabla_{U(T, s)} F(U(T, s))-\Phi(T, s)\right] \delta U^{\dagger}(T,s)\right) &+\operatorname{Tr}\left(\left[\Phi(0, s)\right] \delta U^{\dagger}(0,s)\right)+\\
&+\int_{0}^{T} \operatorname{Tr}\left[\left[\nabla_{U(t, s)} \mathcal{H}+\frac{\partial \Phi(t, s)}{\partial t}\right] \delta U^{\dagger}(t,s)\right]+\left[\nabla_{\varepsilon(t,s)} \mathcal{H}\right] \cdot \delta \varepsilon(t,s) d t
\end{aligned}
\]
\end{widetext}
The corresponding first-order conditions (Euler-Lagrange equations) follow from the requirement that $\delta \mathrm{E}[J]=0$ for any specified deterministic variation $\delta \varepsilon(t,s)$ and hence for any deterministic variation $\delta U(t,s)$.
If the uncertainty is in the system Hamiltonian, $\mathrm{E}\left[\nabla_{\varepsilon(t)} \mathcal{H}\right]=\nabla_{\varepsilon(t)} \mathrm{E}[\mathcal{H}]$.
Then
\begin{widetext}
\[
\small
\begin{array}{l}
\qquad \begin{aligned}
\frac{\partial}{\partial \varepsilon(t)} \mathrm{E}[\mathcal{H}(U, \Phi, \varepsilon)] =& \frac{\partial}{\partial \varepsilon(t)} \sum_{\gamma^{\prime} \in \Gamma}\int_0^{s_f} \left\langle\Phi(t, s),-\frac{i}{\hbar} H(t, s) U(t, s)\right\rangle \exp (-\imath \gamma s) d s\cdot \delta\left(\gamma, \gamma^{\prime}\right) \frac{\mathrm{E}\left[\prod_{k} \theta_{k}^{\alpha_{k}\left(\gamma^{\prime}\right)}\right]}{\prod_{k} \theta_{k}^{\alpha_{k}\left(\gamma^{\prime}\right)}} \\
&=-\frac{i}{\hbar} \sum_{\gamma^{\prime} \in \Gamma}\int_0^{s_f} \operatorname{Tr}\left\{U^{\dagger}(T, s) \nabla_{U} F(U(T, s)) U^{\dagger}(t, s) \mu_{I}(s) U(t, s)\right\} \exp (-\imath \gamma s) d s \times\\
&~~~~~\times \delta\left(\gamma, \gamma^{\prime}\right) \frac{\mathrm{E}\left[\prod_{k} \theta_{k}^{\alpha_{k}\left(\gamma^{\prime}\right)}\right]}{\prod_{k} \theta_{k}^{\alpha_{k}\left(\gamma^{\prime}\right)}}
\end{aligned}
\end{array}
\]
\end{widetext}
where first-order conditions from the expression for $\delta \bar{J}(s)$ have been used to solve for the terminal boundary condition on the costate $\Phi(T, s)=\nabla_{U} F(U(T, s))$ 
and the t,s-evolved costate in the PMP-Hamiltonian function:
\[
\begin{aligned}
&\mathcal{H}(U(t, s), \Phi(t, s), \varepsilon(t)) =\left\langle\Phi(t, s),-\frac{i}{\hbar} H(t, s) U(t, s)\right\rangle \\    
&~~=-\frac{i}{\hbar} \operatorname{Tr}\left\{\Phi^{\dag}(t, s) H(t, s) U(t, s)\right\} \\   
&~~=-\frac{i}{\hbar} \operatorname{Tr}\left\{U^{\dagger}(T, s) \nabla_{U} F(U(T, s)) U^{\dagger}(t, s) \mu(s) \varepsilon(t) U(t, s)\right\}
\end{aligned}
\]
where $\mu$ is also represented in the interaction picture.

The explicit forms of $\nabla_{U} F(U)$ for observables and gates can be substituted above.
For quantum observable control, $\nabla_{U(T)}F(U(T)) = U(T)[U(T)\rho_0 U^{\dag}(T), \Theta] $ \cite{chak2008a}, whereas for gate control,
$\nabla_{U(T)}F(U(T)) = U(T)W^{\dag}U(T)-W$ \cite{moore2011} .
The first-order condition is $\mathrm{E}\left[\frac{\partial}{\partial \varepsilon(t)} \mathcal{H}(U, \Phi, \varepsilon)\right]=0, \forall t \in[0, T]$. Following the principles above, the first-order conditions for any higher moment can be derived, by replacing $F(U(T))$ above
with $F^n(U(T))$. The first-order optimality condition in the presence of field noise can be derived similarly, and will be presented in a future work.
Moreover, as described above in Section \ref{subsetencoding}, identification of quantum pathways that significantly contribute to the quantum control gradient $\frac{\partial}{\partial \varepsilon(t)} \mathrm{E}[\mathcal{H}(U, \Phi, \varepsilon)]$
can be achieved through leading order sensitivity analysis. Here, this involves encoding only those Hamiltonian parameters for which the magnitude of the respect component of $\frac{dJ}{d\theta}$ exceeds a specified threshold value.

Optimization of quantum control moments in the presence of input noise or parameter uncertainty can be achieved using either deterministic or stochastic algorithms. Quantum control Pareto optimization using deterministic, multiobjective gradient-based algorithms has been previously reported \cite{chak2008a, chak2008b}.
One can apply the Pareto optimization methods reported therein to robust quantum control, by replacing the deterministic gradients with the expressions for $\frac{\partial}{\partial \varepsilon(t)} \mathrm{E}[\mathcal{H}(U, \Phi, \varepsilon)]$ derived above for the gradients of quantum control objective moments.
These methods can carry out constrained control optimization wherein one objective can be maximized or minimized while another is held constant.
For example, mean-variance optimization of a single quantum observable can be achieved in this way, especially in model-based control, by optimizing one moment (maximization of $E[J])$ followed by constrained optimization of the second moment (minimization of var$(J)$).
This approach is applicable to any kind of quantum robust control optimization (any moments) for either field noise or Hamiltonian uncertainty.

Gradient-based quantum control optimization algorithms have been applied to optimize quantum control moments using leading order Taylor approximations (for field noise) in \cite{hock2014}.
The quantum robust control methods described herein can be used in conjunction with leading order control methods. Namely, once a relatively robust field is found using leading
order approximations, one can optimize its robustness with more accurate moment calculations. 
Such refinement of the convergence of robust control algorithms that rely on leading order expansions is warranted especially for more demanding quantum control objectives.

Here, we apply stochastic multiobjective algorithms to identify Pareto optima in conjunction with the above robustness analysis techniques for mean-variance optimal control since they are applicable to either model-based or model-free control. However, for model-based control, gradient-based algorithms based on the above expressions can be employed.
We focus on the most common quantum control laboratory objective, namely state to state population transfer.
For any of the Pareto optima identified below, it can be demonstrated that these optima satisfy the robust PMP optimality conditions derived above. Numerical verification of Pareto optimality and the application of gradient-based algorithms will be studied in future work.

\section{Robust Control Optimization via Multiobjective Evolutionary Algorithms}
\label{Robust_Control_via_Evolutionary_Approach}

\subsection{Quantum Learning Control via Genetic Algorithm (GA)}
Optimization of various quantum systems using evolutionary algorithm, such as GA, evolutionary strategies and their variants, has been frequently performed in an experimental setting commonly referred to as the \emph{learning feedback loop} \cite{warr1993}. In the context of process control, this is an example of a model-free control strategy in which the \emph{stochastic} algorithm \emph{learns} the control solution based on trial and error via incoherent feedback measurements. This control implementation inherently takes into account the true system parameters without explicit knowledge of their values such that the issue of the system's model parameter uncertainty is overcome. In this case, the design of the controller is mainly concerned with dealing with input field noise during the optimization. While the laboratory implementation of this control strategy has often been successful, its mechanism of convergence has not been precisely understood. Moreover, in cases where there is significant noise associated with the input field, there is no guarantee that the algorithm would obtain a robust solution. 
In this work, the convergence mechanism of the algorithm is first described in the nominal case and the results are used as the basis for further analysis of model-based and model-free robust quantum control in the next two subsections. Specifically, the convergence of the learning algorithm will be analyzed in terms of addition of quantum pathways and their interferences using the robust control theory developed above.

\begin{figure}
	\centering
	\includegraphics[width=7cm]{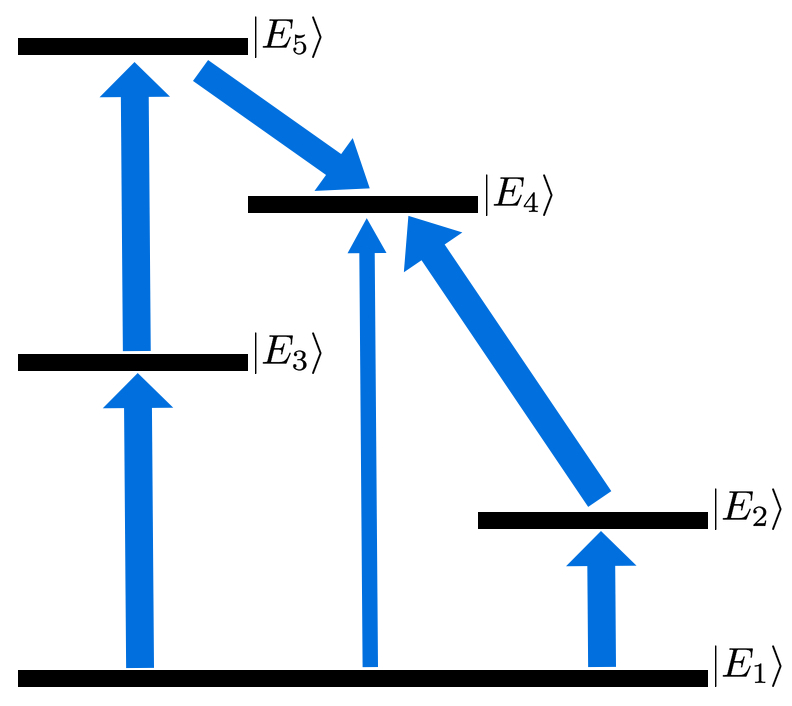}
	\caption{\label{quantum_system}Graphical depiction of a five-level artificial quantum system. The state $|E_4\rangle$ is the target state and it is connected to three states, namely $|E_1\rangle$, $|E_2\rangle$, and $|E_5\rangle$. }
\end{figure}

\begin{figure*}
	\centering
	\includegraphics[width=16cm]{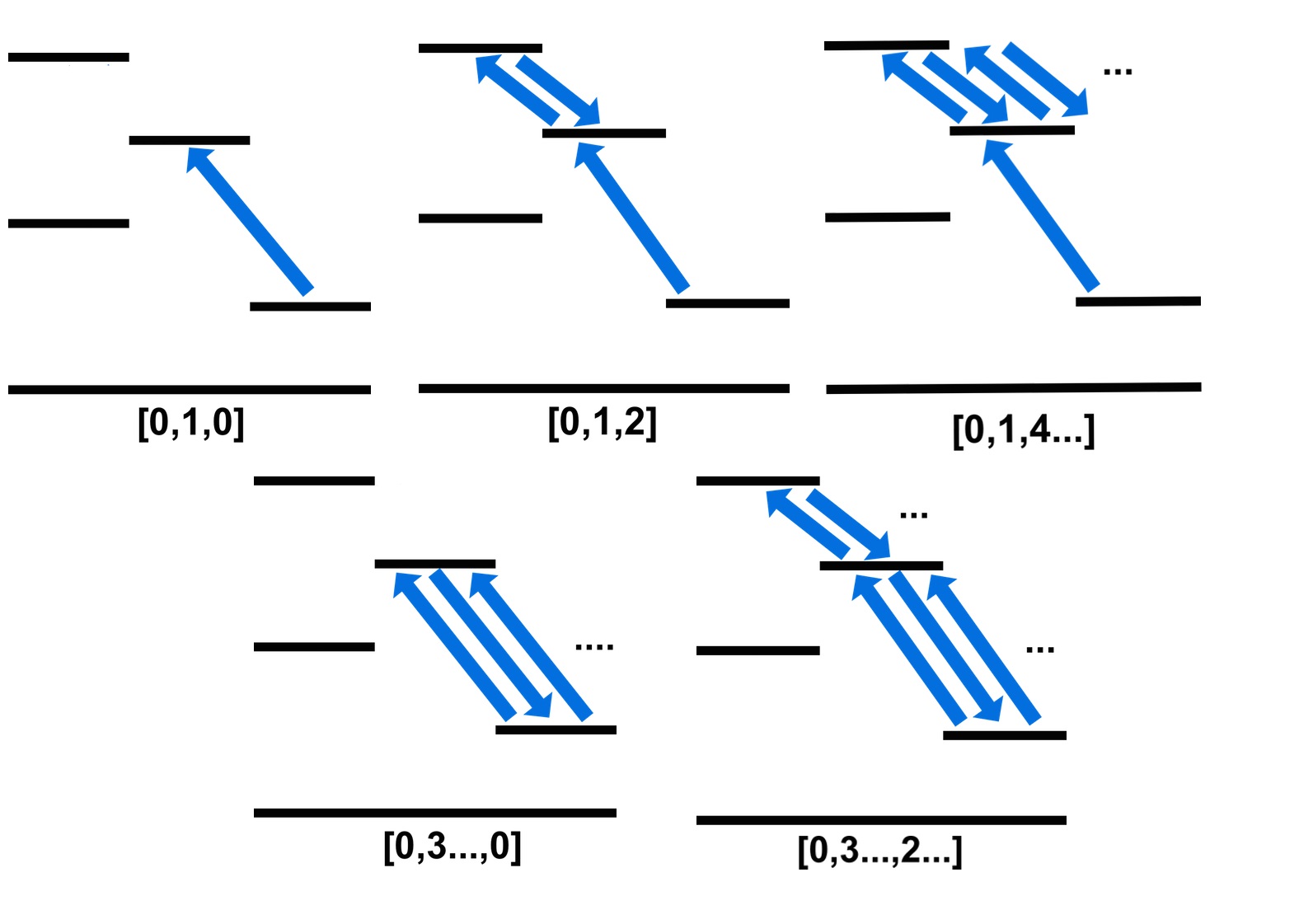}
	\caption{\label{sigpath} Examples of significant pathways in control of the five-level quantum system. The text under each figure represents the polytope $\vec\alpha$ corresponding to each dipole pathway. Note that none of the significant pathways took the direct route of $|E_1\rangle \rightarrow |E_4\rangle$, and is because the magnitude of the dipole moment for the transition is half as strong as the other state-to-state couplings.}
\end{figure*}

The robust control study would be performed in the following artificial quantum system (Figure \ref{quantum_system} and Figure \ref{sigpath}), which is closely related to atomic quantum systems like atomic Rb used in experimental quantum learning control \cite{brif2010,roslundroth2006}: 
\begin{align}
	H_{0}=\begin{pmatrix} 0 & 0 & 0 & 0 & 0 \\ 0 & 0.5 & 0 & 0 & 0 \\ 0 & 0 & 1 & 0 & 0  \\ 0 & 0 & 0 & 1.5 & 0 \\ 0 & 0 & 0 & 0 & 2 \end{pmatrix},~\mu=\begin{pmatrix}0 & 2 & 2 & 1 & 0 \\ 2 & 0 & 0 & 2 & 0 \\ 2 & 0 & 0 & 0 & 2 \\ 1 & 2 & 0 & 0 & 2 \\ 0 & 0 & 2 & 2 & 0 \end{pmatrix}.
\end{align}
%

%

We note that a variety of quantum control systems previously studied, including both molecular and spin states, can be described by Hamiltonians of Hilbert space dimension similar to that above and are amenable to model-based control, rendering the robust control of systems such as our example system important in practice. In the context of atomic and molecular systems, the control of subspectra similar to that above has been studied using the STIRAP method \cite{shapiro1997,kurkal2001,demirplak2006}. More generally, robustness to uncertainty in a small number of Hamiltonian parameters is important in a variety of quantum control applications including quantum information processing, where for example coupling strengths between subsystems may be imprecisely known based on first principles \cite{kosu2013}. Estimation of a limited number of Hamiltonian parameters has been reported in \cite{youngwhaley2009}; the methodologies introduced herein would also be applicable to model-based control of such systems.

The GA chromosome formulation is chosen as $\vec x \equiv [\omega_1, \cdots, \omega_n, \phi(\omega_1), \cdots, \phi(\omega_K)]$. Each $\vec x$ represents an electric field profile $\varepsilon(t)=\sum_{k=1}^K A(\omega_k)\cos(\omega_k t + \phi(\omega_k))$ with a duration $T=40$. The number of modes $K$ is chosen to be 7, and $A(\omega_k)$ 0.15. \\

\begin{figure*}
	\centering
	\includegraphics[width=18cm]{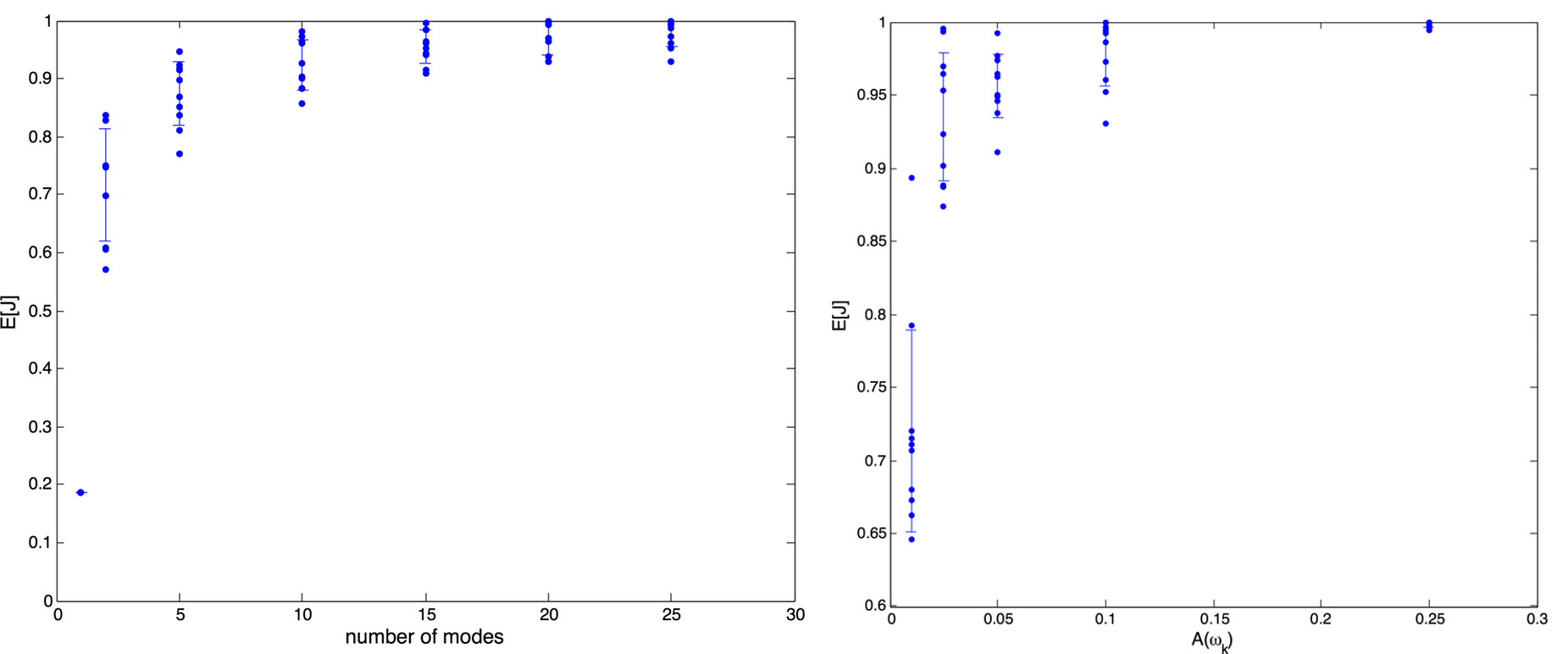}
	\caption{\label{parameffect}}
\end{figure*}

In experimental quantum control, phases, amplitudes, frequencies or a combination thereof can be optimized. Because in this paper, noise in field mode amplitudes and uncertainty in dipole parameters are considered, the parameters for GA optimization were chosen to be frequencies and phases with fixed amplitudes. Otherwise, field noise could cause amplitudes to exceed inequality constraints.  The most natural choice of which parameter set to not sample is the one that is commonly constrained in an experimental setting. Finally, high fidelity control is possible by setting amplitudes for multiple modes to a single value, whereas this is not possible if mode frequencies are constrained in this manner. Note that amplitudes were also sampled across different optimization runs (Figure \ref{parameffect}), 
based on which the value used for Pareto front sampling and robust control was chosen. Similarly, for the number of modes, it was found that the number chosen was sufficient for high fidelity control of this system while not increasing computational expense. Several optimizations with different choices of parameters were studied numerically (Figure \ref{parameffect}). 
%

\subsection{Multiobjective Optimization of Dipole Robustness}
As discussed in the previous section, given uncertain input and system parameters the objective of model-based robust control is to obtain control solutions which satisfy a set of robustness criteria. These criteria include the first and second moment of transition amplitude and probability. In the robust control method proposed, the first step is to perform multiobjective optimization of the two robustness criteria under dipole uncertainty. The amplitude noise would be significant during the experimental implementation of the control, and is studied in the subsequent step.

The analytical expressions for the different moments of transition probability with respect to a control profile have been derived and can be numerically solved. In the examples below, asymptotic robustness analysis has been used alongside numerical sampling from the probability distribution in order to achieve a suitable balance between accuracy and speed. That is, the number of numerical draws from the parameter distribution (e.g. the Gaussian distribution associated with uncertain dipole parameters) that is required to determine each quantum control moment to a specified level of accuracy was determined periodically during the course of the optimization using asymptotic robustness analysis. This is required to guarantee the algorithm is in fact converging to the true Pareto optima. Depending on the complexity of the quantum system, dimensionality reduction using significant pathways identification can be used to further mitigate computational expense. 

Depending on user preferences for maximization of expected fidelity or minimization of variance the appropriate point on the front can be defined as desired control. The solutions derived in this optimization can then serve as initial population for learning control of amplitude robustness described in the next subsection. In addition, the resulting solutions are further analyzed using the asymptotic robustness analysis method for insights into robustness mechanism, which is essential in generalizing the observed results to general quantum systems.

\subsection{Learning Control of Amplitude Robustness}
In addition to robust optimization under dipole parameter uncertainty, robust control of quantum systems in the presence of field amplitude noise has also been studied in the past \cite{akos2014,hock2014}. Significant amplitude noise arises in quantum control experiments involving intense ultrafast lasers in a feedback loop. In the previous section, the robust control objective is the expected transition probability, which is estimated over a number of measurements, and is an optimization over a static landscape. In this case, a traditional GA (tGA) has frequently been shown to successfully optimize different quantum systems \cite{goldberg1989,rabi2003}.
However, there are cases where the optimization must be performed without averaging when the optimization would take place across a noisy landscape and the locations of optima vary from one optimization iteration to the next. In this case, a GA which has not been appropriately equipped with operators to \emph{explore} and \emph{exploit} the landscape would not converge to a diverse population of robust solutions. In this work, studies on how robust control may be achieved in this dynamic setting are further described and compared to its expected formulation counterpart. Dynamic optimization using tGA often fails due to loss of diversity. This takes place because of static selection pressure, whereby an elite individual in the population dominates the reproductive pool by replacing outlier individuals after one particular iteration. Another cause for population uniformity in tGA is weak exploration feature. Here, diversity lost through selection and crossover is not recovered via mutation, resulting in population inbreeding \cite{Mcgi2011}. Thus, \emph{adaptive} mutation and crossover rates and selection pressure are essential to preserve the ability to explore other parts of the landscape as the landscape fluctuates due to noise. ACROMUSE (Adaptive CROssover, MUtation, and SElection) GA \cite{Mcgi2011} has recently been developed and applied to multi-modal test functions using binary-coded GA formulation (Algorithm \ref{algo:acromuse}). In this study, the algorithm is applied using a real-coded formulation \cite{deb1995} to robust dynamic quantum control in the presence of amplitude noise.

\begin{algorithm}[H]
\caption{ACROMUSE Pseudocode}
\label{algo:acromuse}
	\begin{algorithmic}[2]
		\STATE $\mathrm{X} \leftarrow init()$
		\STATE $\mathrm{f} \leftarrow eval(\mathrm{X})$
		\STATE $[\mathrm{HPD},\mathrm{SPD}] \leftarrow diverse(\mathrm{X}) $: measure diversity in terms of \textit{HPD} and \textit{SPD}
		\STATE $[\mathrm{p_c},\mathrm{p_m}] \leftarrow prob(\mathrm{HPD},\mathrm{SPD})$: determine crossover $\mathrm{p_c}$ and mutation probability $\mathrm{p_m}$
		\STATE $\mathrm{R} \leftarrow select(\mathrm{X},\mathrm{f},\mathrm{HPD},\mathrm{SPD})$: select fit individuals based on $\mathrm{HPD}$ and $\mathrm{SPD}$
		\WHILE{$t < t_{max}$}
			\STATE \hspace{0.2in} $\mathrm{C}\leftarrow recomb(R,p_c)$
			\STATE \hspace{0.2in} $\mathrm{M} \leftarrow mutate(R,p_m)$
			\STATE \hspace{0.2in} $\mathrm{f} \leftarrow eval(\mathrm{C},\mathrm{M})$	
			\STATE \hspace{0.2in} $[\mathrm{p_c},\mathrm{p_m}] \leftarrow prob(\mathrm{HPD},\mathrm{SPD})$
			\STATE \hspace{0.2in} $[\mathrm{HPD},\mathrm{SPD}] \leftarrow diverse(\mathrm{X})$		
			\STATE \hspace{0.2in} $\mathrm{R} \leftarrow select(\mathrm{X},\mathrm{f},\mathrm{HPD},\mathrm{SPD})$
			\STATE \hspace{0.2in} $t \leftarrow t+1$
		\ENDWHILE
	\end{algorithmic}
\end{algorithm}
In addition to adaptive operators, population diversity may be achieved using other techniques including various \textit{niching} methods such as fitness sharing (FS) and deterministic crowding (DC). While the \emph{No Free Lunch} theorem states that any "two genetic algorithms are equivalent when their performance is averaged across all possible problems", ACROMUSE GA is specifically suited to the quantum control problem as it adapts quickly to fitness landscape variation and to the exploration of highly multimodal quantum control landscapes. It is also important to note that there is implicit averaging associated with general GA optimization; a GA of infinite population size, which uses proportional selection and adds random perturbations to the design variables in each generation, is theoretically equivalent to GA optimization of the expected fitness function \cite{Mcgi2011}. Thus, the performance of tGA and ACROMUSE GA are performed using the same population size.
\begin{figure*}
\includegraphics[width=18cm]{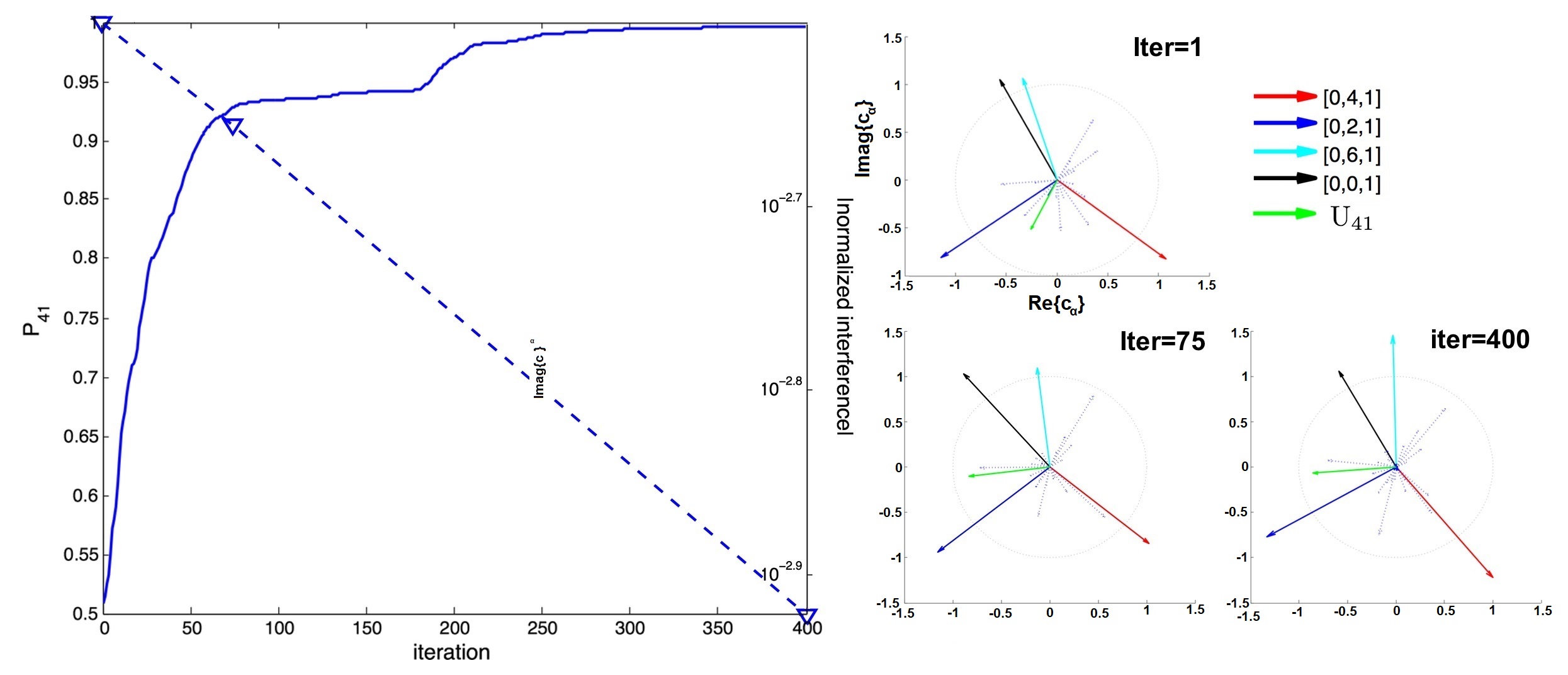}
\caption{\label{opttraj}  GA optimization trajectory of nominal $P_{41}$ (solid) and its normalized interference (dashed) (left) with identified significant pathways (right). As described in the text, the total interference is negative (destructive) and its magnitude decreases due to optimization.  The plot shows how the magnitude of pathways and their relative angles evolve during the optimization.}
\end{figure*}
\begin{figure}
	\includegraphics[width=9cm]{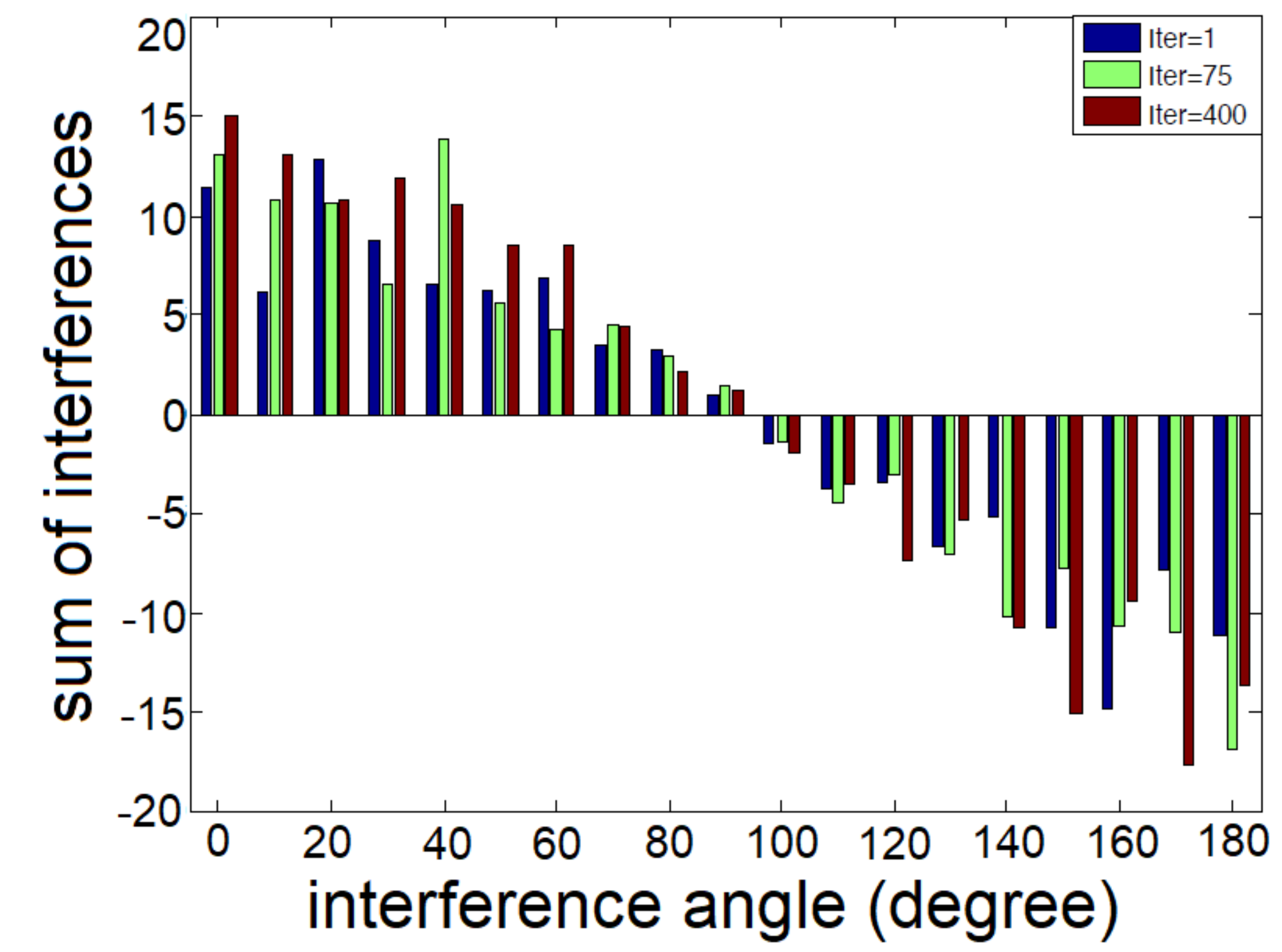}
	\caption{\label{suminterference} Bar plot of sum of interferences along the GA optimization trajectory. As descri bed in the text, the destructive interference (negative bars) is minimized relative to the constructive ones as the GA converges to an optimal solution; the total interference is negative (destructive) throughout the course of optimization. } 
\end{figure}

\section{Results and Discussions}
\label{Results_and_Discussions}
\subsection{Quantum Optimal Control via GA}
The result of the optimization is analyzed using the subset encoding of dipole pathways. The encoding is performed over the algorithmic iterations to track how the quantum pathways and their interferences evolve over the course of the optimization. Some examples of transition pathways for the optimal field under nominal conditions are graphically depicted in Figure \ref{sigpath}. 
As seen in Figure \ref{opttraj}, as GA learns the direction of optimal control under nominal condition, the pathway amplitudes increase and their interferences become stronger. As shown analytically in \cite{mitr2003} and \cite{akos2014}, the pathway amplitude is always positive and can reach values above unity. Consequently, the interference is always destructive in the case of optimal control. Furthermore, over the course of the optimization, the algorithm is shown to minimize the \emph{normalized} destructive interference (Figures \ref{opttraj},\ref{suminterference}). It is also worthwhile to note that different GA runs converge to different solutions due to the level sets 
of quantum control landscapes \cite{chak2007}. In addition, fields which are limited in their parametrization lead to non-optimal landscapes and therefore suboptimal solutions (Figure \ref{parameffect}). 
Analysis of suboptimal control solutions reveals that, in comparison, a sub-optimal field utilizes fewer pathways such that pathway amplitude and interferences are not maximized for maximum transition probability (Figure \ref{opttraj} and Table \ref{RobustPath}). However, as will we see below, more pathways mean more entry points for noise and such robust control does not necessarily mean optimal control under nominal condition.

\begin{figure*}
	\includegraphics[width=18cm]{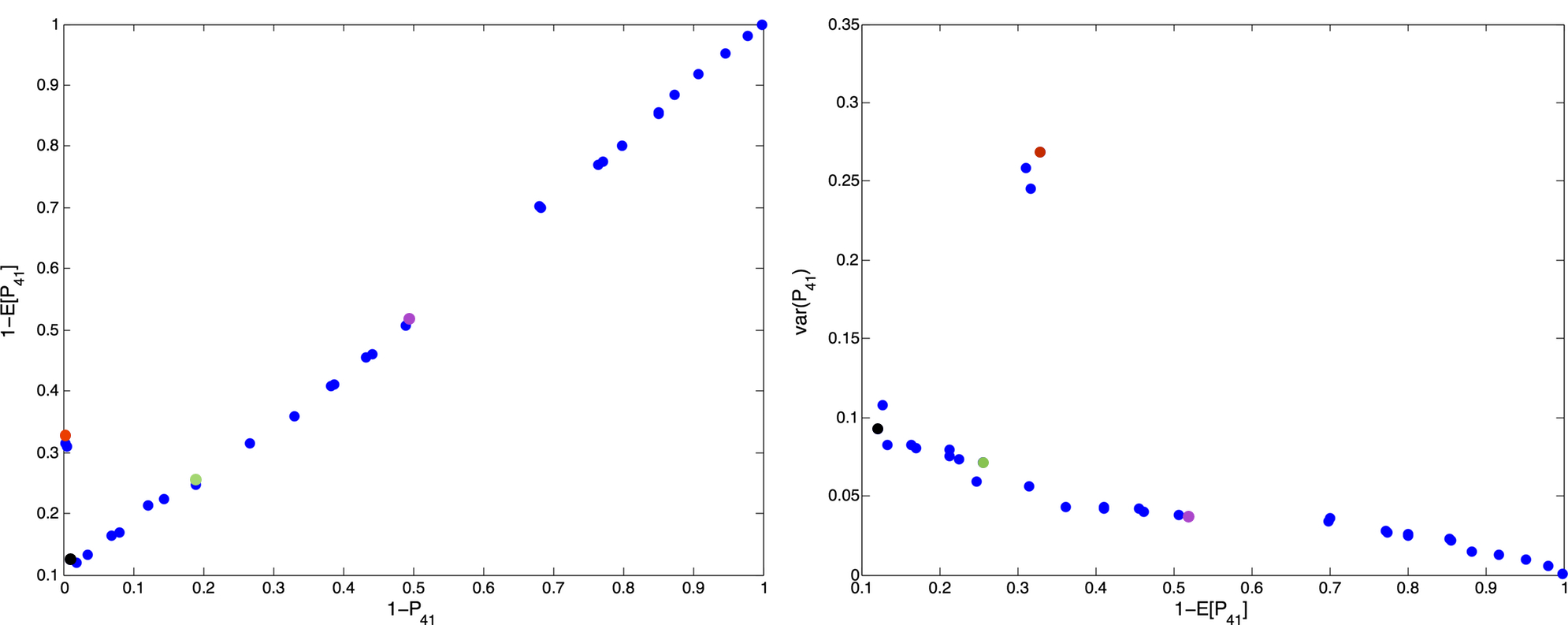}
	\caption{\label{paretosols1}Pareto front of robust control solutions in terms of the first and second moment of the transition probability. The plot shows that the optimal solution in the nominal case is not the most robust in the Pareto sense, i.e. it resides in the non-dominated front of the E[$P_{41}$] and var($P_{41}$) Pareto surface.} 
\end{figure*}

\subsection{Multiobjective Optimization of Dipole Robustness}
\begin{figure}
	\includegraphics[width=9cm]{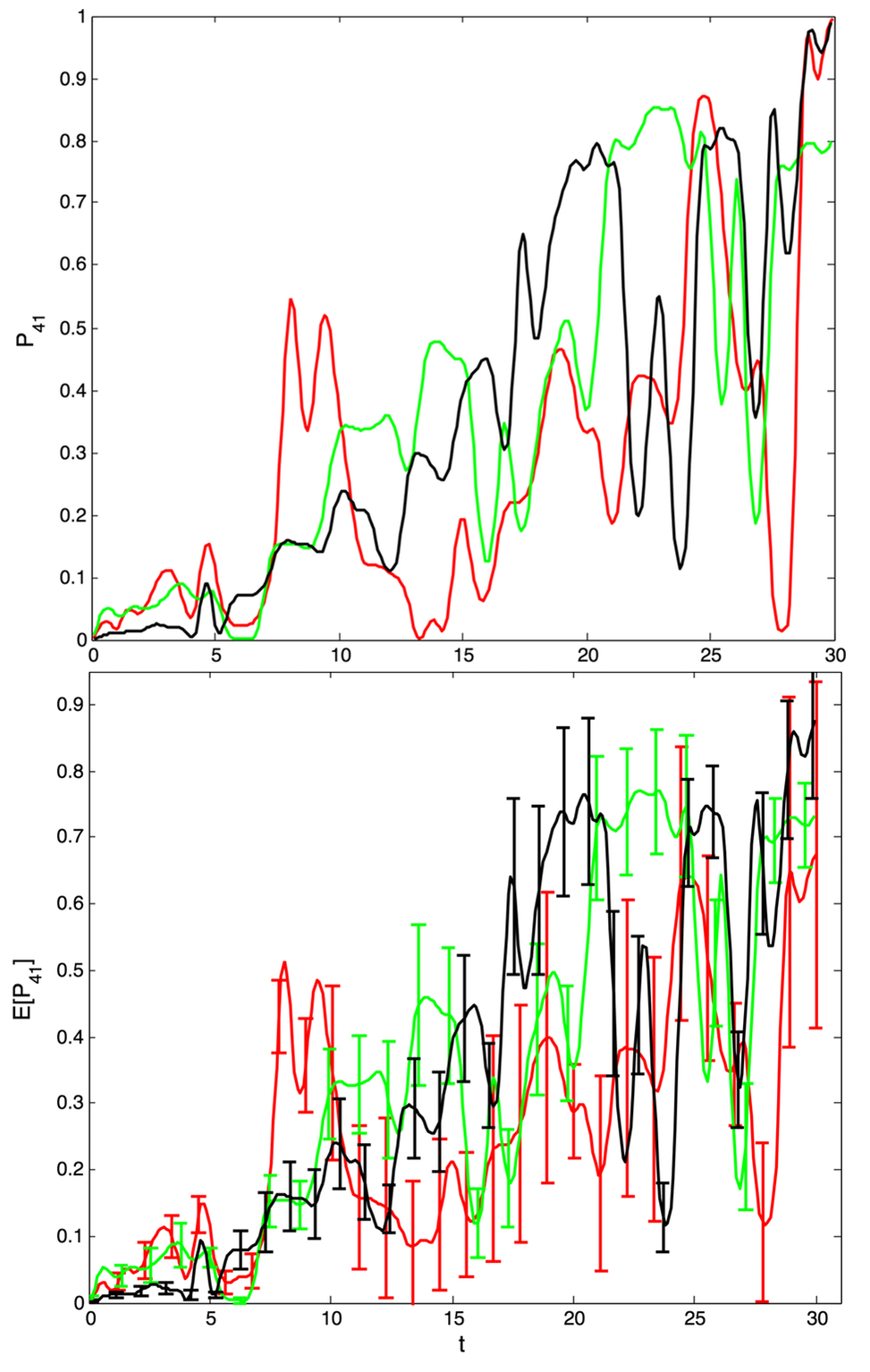}
	\caption{\label{paretosols2} Plot of population transfer of Pareto front solutions under nominal (top) and uncertain conditions (bottom),
for the color-coded fields depicted in Figure \ref{paretosols3}.
The plots show a robust solution may be non-optimal under nominal conditions and a nominally optimal solution may not be robust under uncertain conditions.}
\end{figure}
\begin{figure}
	\includegraphics[width=9cm]{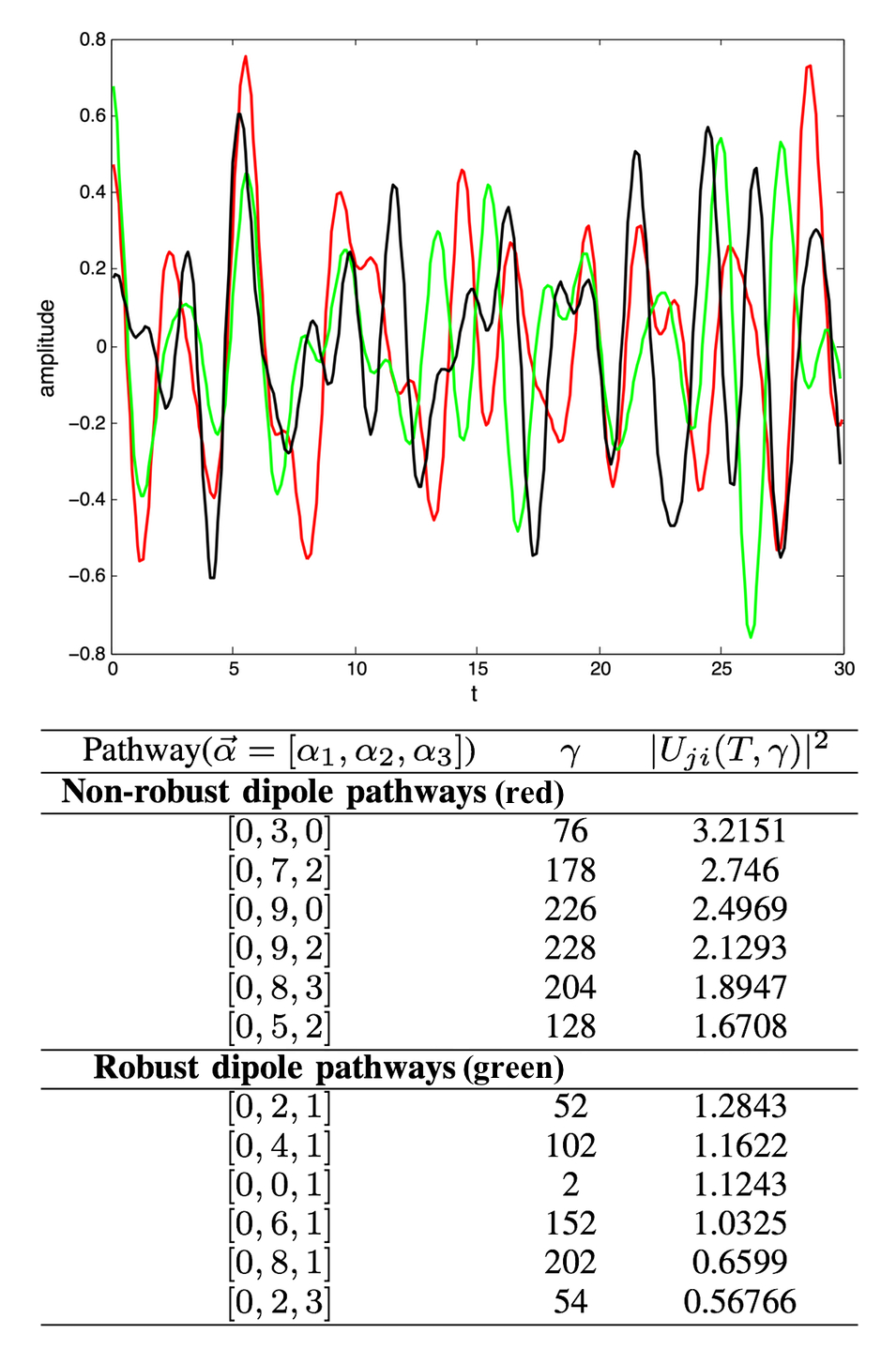}
	\caption{\label{paretosols3} Plot of 
fields corresponding to the color-coded Pareto front solutions depicted in Figure \ref{paretosols1}, and a list of associated pathways for robust (lowest value of var($P_{ji}$)) and non-robust fields. It is observed that the non-robust dipole pathways 
utilize higher order pathways relative to their robust counterparts.}
\end{figure}

NSGA-II \cite{Deb2002} is used to obtain the 3D Pareto front in terms of the nominal, expected and variance of the transition probability (Figure \ref{paretosols1}). The 3D Pareto front can be presented more clearly upon the front projection towards the 2D plane of $J_{nom}$/E[$J$] front and E[$J$] /var($J$) front (Figure  \ref{paretosols1}). Figure \ref{paretosols1} suggests that there exists robust solution which is close-to but not optimal under nominal conditions. Specifically, there exists a robustness trade-off in terms of E$[J]$ and var($J$), which include non-optimal solutions shown as dominated points in the Figure. Another observation can be made in that non-optimal pulse which does not induce optimal transition from an initial to a target state is also less susceptible to noise. This makes intuitive sense as the quantum pathways resulting in the state-to-state transitions are low in amplitude such that uncertainty affecting those pathways do not manifest in affecting the population transfer. In order to generalize the observation for understanding robust control of general quantum systems, analysis of the Pareto fronts is performed using the subset of pathway encoding. The analysis reveals the mechanism of dipole robustness in terms of dipole pathways utilized by the control field. Figures \ref{paretosols2} and \ref{paretosols3} display the real-time nominal and expected transition probabilities as well as real-time field amplitudes for solutions on the Pareto front, showing how a robust solution may be nonoptimal under nominal conditions.
\begin{table*}
\caption{\label{RobustPath}List of robust pathways and their magnitudes corresponding to different transition probability under expected and nominal conditions shown in Figure \ref{paretosols1}.}
\centering
\small
\begin{tabular}{cc|cc|cc}
	\hline
	\textbf{Near-optimal}		&\textbf{magnitude} 	&\textbf{Sub-optimal}	&\textbf{magnitude}	&\textbf{Non-optimal}	&\textbf{magnitude}\\
	\textbf{(black)}	&			&\textbf{(green)}	&			&\textbf{(red)}		&		\\\hline
 	$[0, 4, 1]$		&2.7460		&$[0, 2, 1]$	&1.2843		&$[0, 4, 1]$	&1.0548   		\\
	$[0, 5, 0]$		&2.4865		&$[0, 4, 1]$	&1.1622		&$[0, 2, 1]$	&1.0128		\\
	$[0, 7, 0]$		&2.1538		&$[0, 0, 1]$	&1.1243		&$[0, 0, 1]$	&0.9775		\\
	$[0, 6, 1]$		&1.8544		&$[0, 6, 1]$	&1.0325		&$[0, 6, 1]$	&0.9444		\\ 	
	$[0, 2, 1]$		&1.8532 		&$[0, 8, 1]$	&0.6599		&$[0, 8, 1]$	&0.5883		\\
 	$[0, 3, 0]$		&1.2933		&$[0, 2, 3]$	&0.5677 		&$[0, 2, 3]$	&0.4760		\\	
	$[0, 9, 0]$		&1.0531		&$[0, 0, 3]$	&0.4363		&$[0, 0, 3]$	&0.3737		\\ 	
	$[0, 4, 0]$		&0.8428		&$[0, 4, 3]$	&0.4323		&$[0, 5, 0]$	&0.3566		\\
 	$[0, 4, 3]$		&0.7860		&$[0, 7, 0]$	&0.2900		&$[0, 4, 3]$ 	&0.3332		\\
	$[0, 0, 1]$ 	&0.7275		&$[0, 5, 0]$	& 0.2768		&$[0, 7, 0]$ 	&0.2813	 	\\
\end{tabular}
\end{table*}
\begin{table}
\caption{\label{AcromuseSols}List of ACROMUSE solutions which use different pathways and are affected by noise in different ways and at different instances.}
\centering
\begin{tabular}{|c|c|c|c|}
	\hline
	$\vec x_1$		&$\vec x_2$ 	&$\vec x_3$	&$\vec x_4$	\\ \hline
	1.7384		&1.7429		&1.7399		&3.2723		\\	
    	1.0448		&3.9752		&1.0452		&3.9819		\\
   	1.3715		&1.3703		&1.3665		&2.7221		\\
    	1.3669		& 1.3627		&1.3660		&3.9054		\\
   	1.5256		&1.5262		&1.5250		&0.7801		\\
   	0.7038		&0.7037		&0.7036		&2.0866		\\
  	3.2220		&3.2204		&3.2292		&1.3671		\\
  	2.4826		&2.4794		&0.1077		&5.0870		\\
   	1.9833		&1.9684		&3.9931		&2.4549		\\
  	0.2558		&0.2535		&4.5513		&4.4031		\\
  	6.1403		&6.1408		&6.1403		&0.7070		\\
 	5.5389		&5.5248		&5.5380		&4.3336		\\
    	1.6142		&1.6178		&1.6117		&5.4109		\\
   	4.9542		&4.9673		&4.9537		&1.3264		\\ \hline
\end{tabular}
\end{table}
\begin{table}
\caption{\label{AcromuseCond}List of real-time transition probability associated with ACROMUSE solutions under different noisy instances.}
\centering
\begin{tabular}{|c|c|c|c|}
	\hline
	instance 1			&instance 2 		&instance 3		&instance 4	\\ \hline
	\textbf{0.8556}		&0.6204			&0.8528			&0.4272		\\	
    	0.8118			&\textbf{0.9319	}	&0.8570			&0.3289		\\
   	0.6263			&0.4128			&\textbf{0.8598}	&0.1700		\\
    	0.3380			&0.6006			&0.8332			&\textbf{0.7671} \\ \hline
\end{tabular}
\end{table}
%
\begin{table}
\caption{\label{AcromuseNoise}List of real-time amplitude (including noise) associated with different solutions' modes in \ref{AcromuseSols}
and noisy real-time transition probability in Table \ref{AcromuseCond}.}
\centering
\begin{tabular}{|c|c|c|c|}
	\hline
	amplitude 1			&amplitude 2 		&amplitude 3		&amplitude 4	\\ \hline
	0.1396	&0.1465		&0.1865		&0.1144		\\
	0.1092    	&0.2271		&0.0757 		&0.1569		\\
	0.1436    	&0.0945		&0.1097		&0.1625 		\\
	0.1407    	&0.1452 		&0.1523		&0.1663 		\\
	0.1024    	&0.1425 		&0.2023 		&0.1698  		\\
	0.1942    	&0.1701 		&0.1629		&0.2031		\\
	0.1223	&0.1577		&0.1756		&0.1728		\\ \hline
\end{tabular}
\end{table}
%

\begin{figure*}
	\includegraphics[width=18cm]{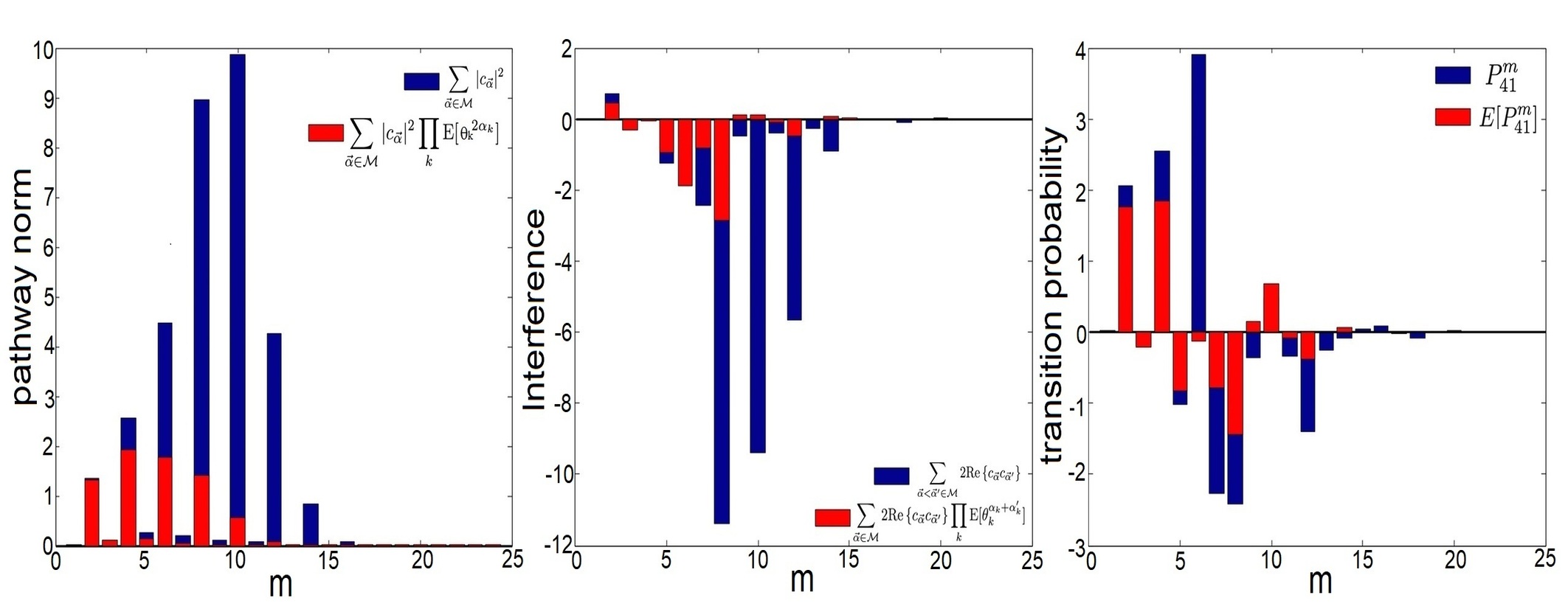}
	\includegraphics[width=18cm]{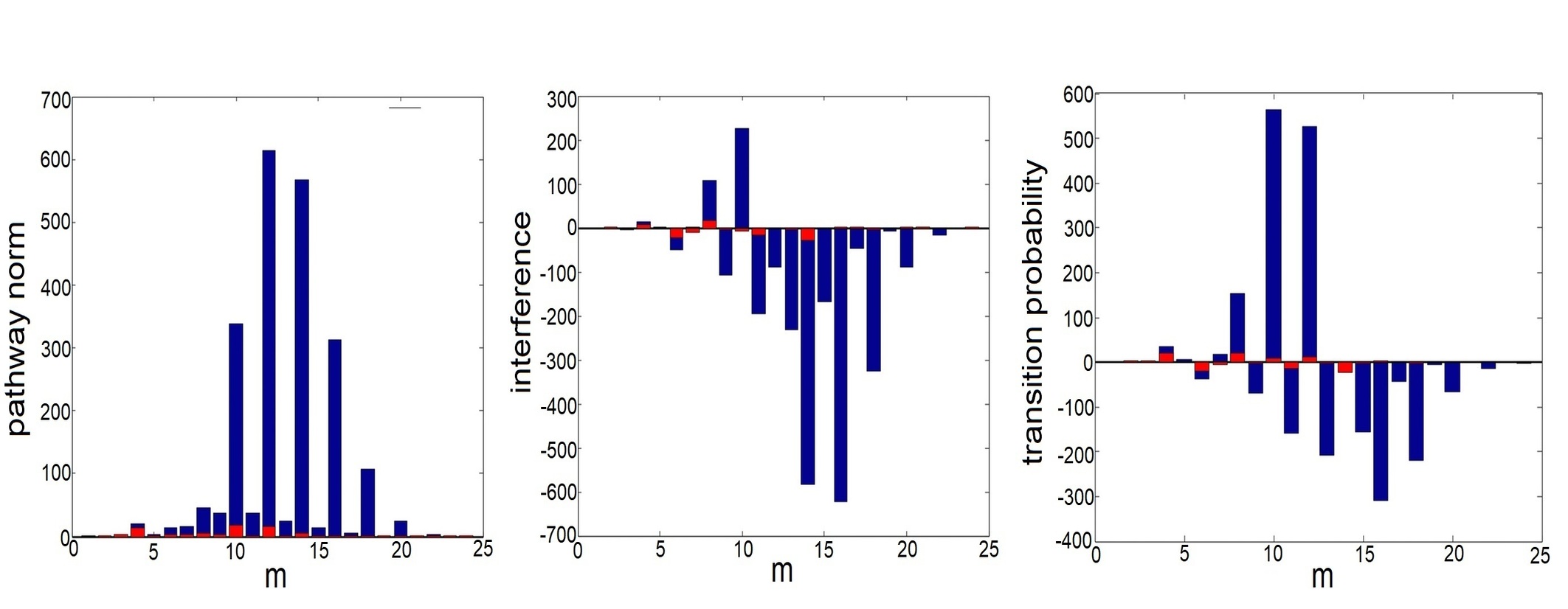}
	\caption{\label{robustMI}  Robustness analysis of a robust Pareto front solution (high E[$P_{ji}$], low var($P_{ji}$)) (\textbf{top}) and  a PF solution with low E[$P_{ji}$], low var($P_{ji}$)  (\textbf{bottom}) in Figure \ref{paretosols1} under Hamiltonian uncertainty via subset encoding. In each case, the first two panels compare the components of each encoded pathway with and without the terms that are affected by uncertainty. The first panel depicts pathway norms whereas the second depicts quantum pathway interferences. The third panel compares the magnitudes of each pathway in expectation and under nominal condition. The optimal solution uses a significantly larger number of higher order pathways compared to the robust counterpart, since the robust control optimization selects pathways that employ lower orders which are less susceptible to uncertainty. Moreover, the mechanism of robustness is such that there is an appropriate number of higher-order pathways utilized for reaching an optimal balance between the different moments of transition probability.} 
\end{figure*}

\begin{figure}
	\includegraphics[width=9cm]{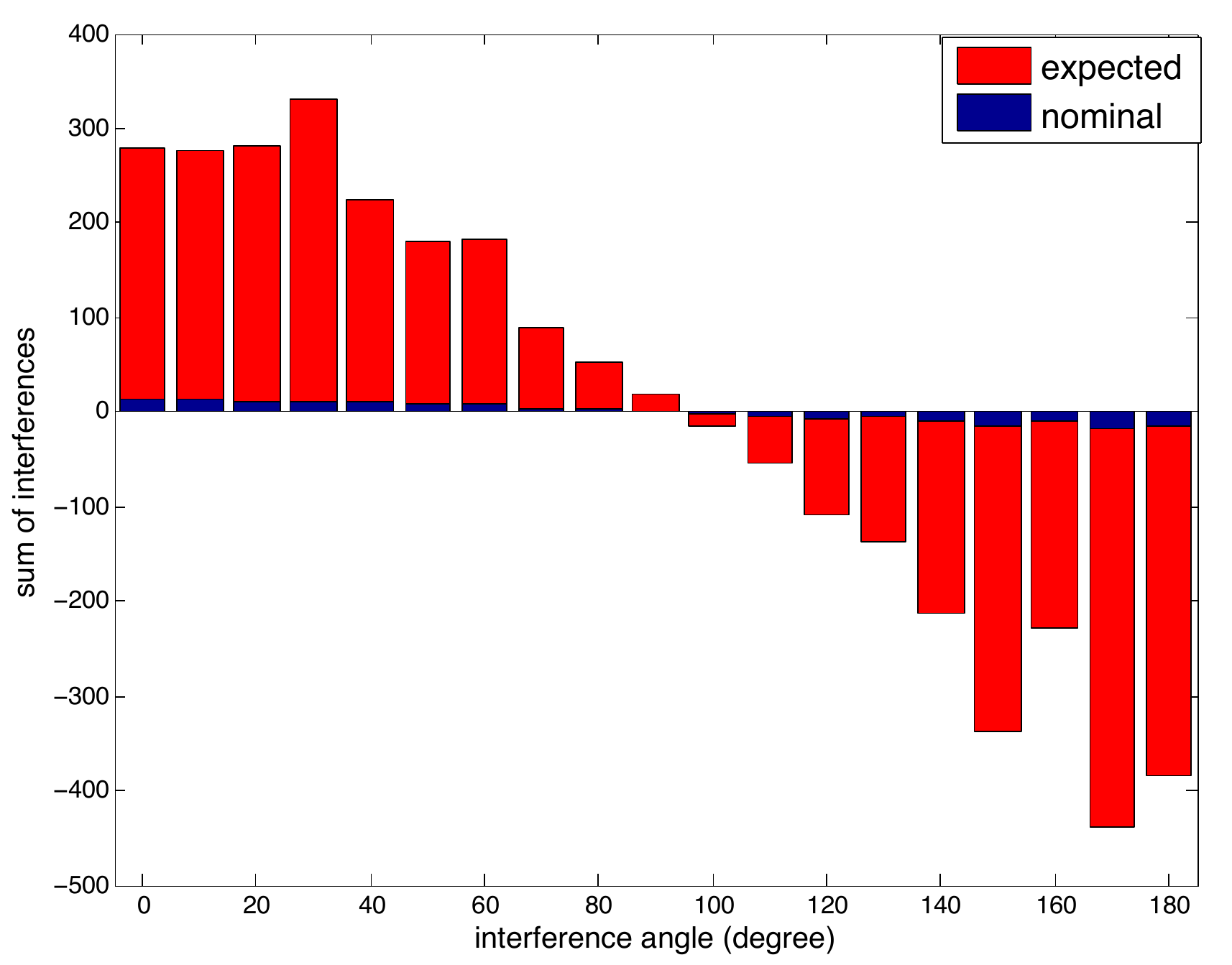}
	\caption{\label{barplot} Bar plot of expected and nominal interferences. The figure indicates how noise increases the average interferences between different pathways such that the destructive interference is amplified relative to the constructive counterpart.} 
\end{figure}

\begin{figure*}
	\includegraphics[width=18cm]{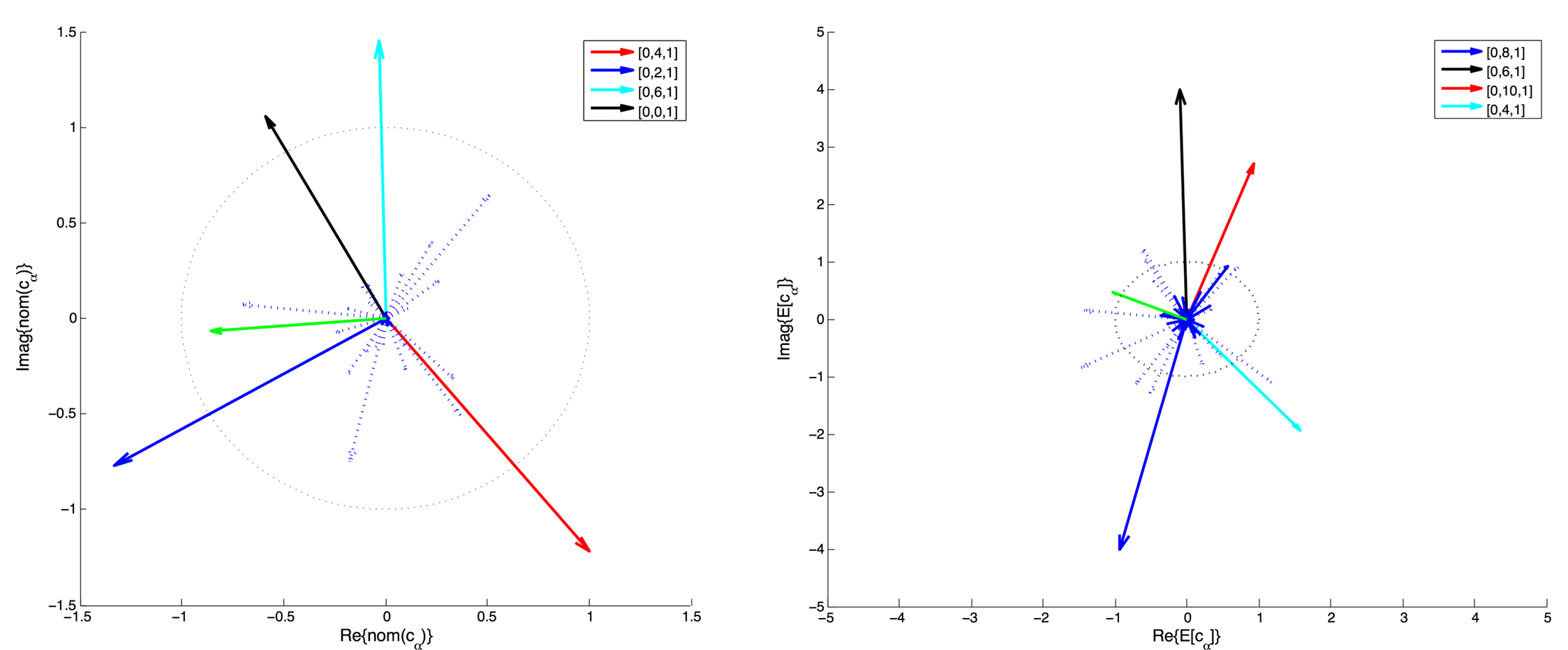}
	\caption{\label{nominal_vs_expected_pathways} Complex plane representation of pathways under nominal (left) and expected conditions with normal dipole uncertainty distribution (right). The figures show how noise significantly change the norm and interference of different pathways resulting in further destructive interference.} 
\end{figure*}

Figure \ref{robustMI} 
present robustness analysis of PF solutions (high E[$P_{ji}$], low var($P_{ji})$ and low E[$P_{ji}$],low var($P_{ji})$ points), decomposing the pathway contributions into the deterministic and stochastic contributions. The analysis reveals how the robust control optimization selects pathways that are less susceptible to the effects of uncertainty. While the optimal solution under nominal condition utilizes the largest number of quantum pathways, including higher-order pathways, in order to satisfy the control objective, the robust solution is one which utilizes the appropriate number of pathways to satisfy the expected transition probability without involving too many of them and higher order ones such that the control is susceptible to noise. 
Furthermore, gleaning from the results of robustness analysis in Table \ref{RobustPath}, there may be a different conditions of robustness relevant in other quantum systems. For instance, there could be circumstances where the interaction between the control field and the dipole parameter is not sufficiently strong to result in optimal combination of pathways, but are associated with a sufficiently small uncertainty such that its expected value and variance is robust. 
Understanding the mechanisms by which robust control fields achieve robustness also requires analysis of the robustness of quantum interferences to uncertainty. First, note that uncertainty amplifies the magnitude of nearly all quantum interferences in expectation, as shown in Figure \ref{barplot},
which represents quantum interferences binned by interference angle. 
The polar representation of interferences bar plots in this Figure have consistently higher total interference magnitudes compared to nominal in each bin - i.e., the various interference angles magnify the effects of uncertainty to a similar extent.
The Figure also indicates how uncertainty increases the average interferences between different pathways such that the overall destructive interference is amplified relative to the constructive counterpart. Second, as noted above in Section \ref{Polar_Representation_of_Quantum_Interferences}, there is a difference between expectation of interferences and the dot product of expectation of pathways. Thus, complex plane representations of dominant pathway contributions to population transfer by nominally optimal vs robust control fields (see Figure\ref{nominal_vs_expected_pathways}) 
cannot fully capture the mechanisms by which robust control fields achieve robustness. 
The expectation of interferences are displayed instead by pathway order in Figure \ref{robustMI} for a robust field and 
for a field with low $E[P_{ji}]$ and low $\mathrm{var}(P_{ji})$. Finally, the order $m$ in partial encoding (Section \ref{subsetencoding}) is the relevant quantity when considering the effect of noise in magnifying interferences, since only those parameters contributing to m are subject to uncertainty.
Recall that $m$ in partial encoding does not refer to Dyson series order.

In summary, while the ability to control a multitude of pathways and their interferences is essential for achieving robust quantum control fidelity, the optimal strategies are quite different for deterministic and robust control. This has been demonstrated numerically by how a nominally optimal control strategy can be very different from a robust optimal control strategy in terms of the pathways and interferences it must control. Neither the deterministic theory of optimal control nor leading order approaches to quantum robust control establishes this result.

%


\begin{figure*}
	\includegraphics[width=18cm]{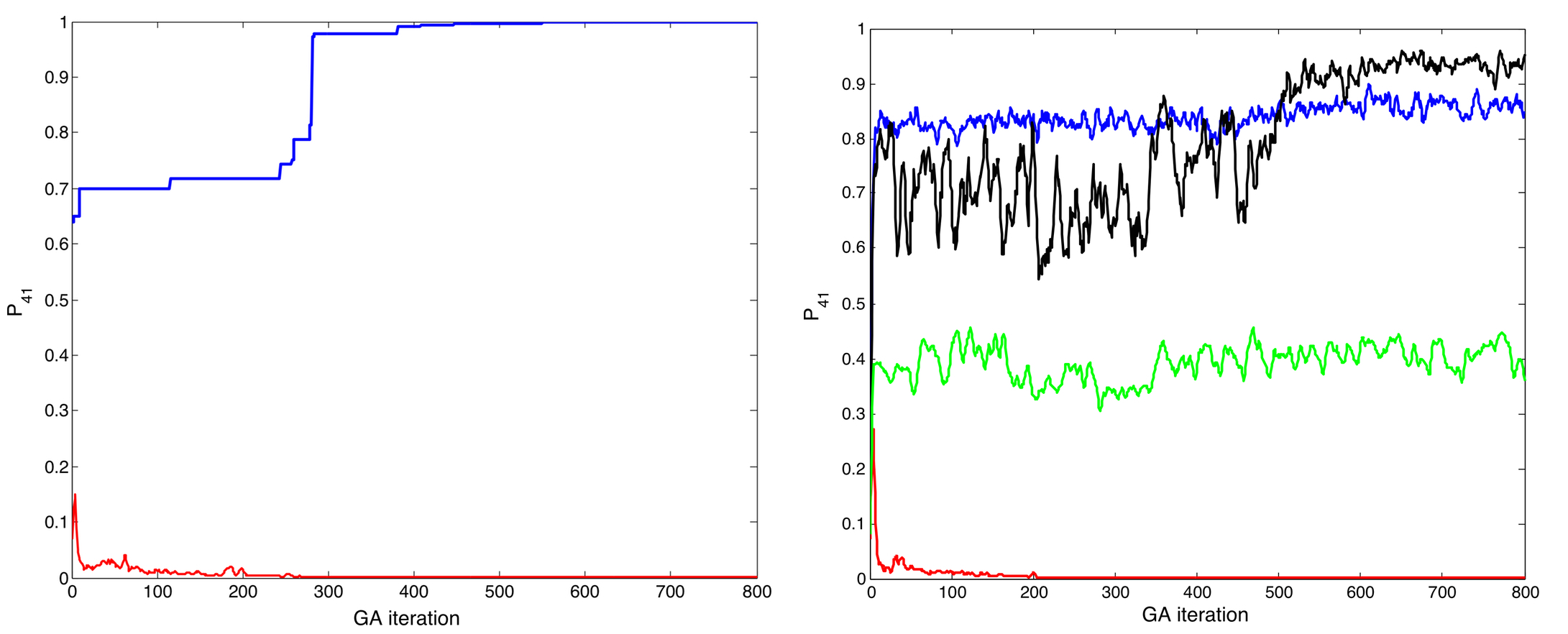}
	\caption{\label{diversoptim} The trajectory of GA optimization under nominal condition (left) and in the presence of amplitude noise (right). The blue line corresponds to the GA trajectory with fixed crossover and mutation operator, where the black line corresponds to the adaptive counterpart. The red (dark gray) and green (light gray) lines are the measure of the HPD along the optimization trajectory for the fixed and adaptive crossover and mutation rates, respectively.}
\end{figure*}
%
%

\subsection{Learning Control of Amplitude Robustness}
\begin{figure*}
	\includegraphics[width=18cm]{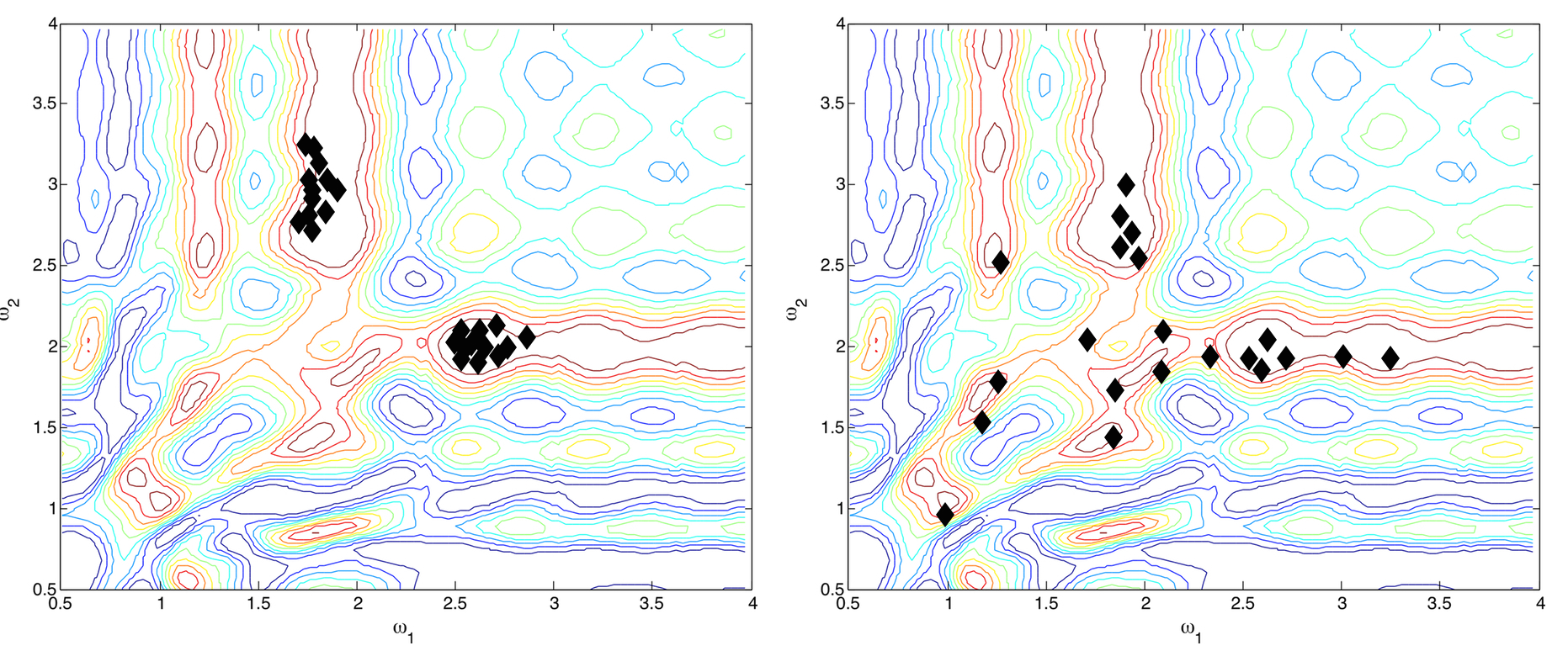}
	\caption{\label{diverslandsc} Schematic of the 2D landscape of tGA (left) and of ACROMUSE optimization (right). The contour plot shows the low distribution of tGA solutions due to the fixed crossover and mutation rates when compared with that of ACROMUSE solutions which employes adaptive crossover and mutation rates.}
\end{figure*}

ACROMUSE and tGA are used to optimize for noisy transition probability (without averaging) and expected transition probability, respectively. The purpose of this simulation is to demonstrate different methods of model-free learning control which can be implemented in a feedback loop. 
Solutions and their properties are listed in Tables \ref{AcromuseSols},  \ref{AcromuseCond} and  \ref{AcromuseNoise}.
The result shows that while GA optimization can be readily performed under static landscape without a diverse population (Figure \ref{diverslandsc} (left) and \ref{diversoptim} 
(left)), the same algorithm fails when dealing with a noisy landscape, where the global optima fluctuate
(\ref{diversoptim} 
(right)).
In the case of the latter, a diverse population must be maintained during the optimization and this is achieved using ACROMUSE GA, with the parameters tuned to ensure exploration and exploitation (Figure \ref{diverslandsc} (right)). Note that the mechanism of convergence is different in the two methods. In tGA optimization of expected transition probability, the algorithm converges to a particular solution in the averaged and static landscape. In contrast, the noisy optimization using adaptive GA results in a distribution of solutions some of which have higher transition probability relative to the others under an instance of input parameter, which changes from one iteration to the next.

\subsection{Integrated Model-Based and Model-Free Control}

The results for multiobjective optimization of dipole robustness can be placed in the learning control counterpart for integrated model-based and model-free quantum robust control. The performance of such a strategy depends upon the magnitude of uncertainty and noise associated with the system and input parameters, respectively. The two types of robust control considered above, namely robust control under system parameter uncertainty and input noise, were chosen because of their potential for combination in the context of an integrated model-based and model-free control strategy. Thus far, there have been very few if any examples of model-based quantum control used to control population transfer in molecules. The principle is to use the uncertain Hamiltonian to do model-based robust control based on moments, then to refine the solutions in the presence of field noise with the  true known Hamiltonian in a model-free learning control approach. 
Moreover, the integrated strategy is an example of adaptive feedback control, wherein the Hamiltonian parameter estimates obtained through asymptotically efficient estimators \cite{chak2012}  can be updated following learning control, after which model-based robust control can again be applied in order to improve convergence. Thus far, although learning control has been implemented in the quantum control laboratory, adaptive feedback control has not.
%
\newpage
\section{Summary and Outlook}
\label{Summary_and_Outlook}
We have described model-based and model-free robust control strategies and explained their methods of convergence using the asymptotic robustness analysis method. The analysis demonstrates the mechanism of how these classes of control methods converge to a particular set of robust solutions. The more restricted the manipulable variables and the larger the variations associated with the system and input parameters are, the harder it is for an optimizer to reach a robust control solution.
In model-based robust control, a multi-objective genetic algorithm is used to obtain Pareto optimal solutions in the presence of Hamiltonian uncertainty, in terms of the nominal and the expected transition probability as well as with respect to the expectation and variance of the transition probability. Analysis of robustness of Pareto optimal solutions showed that the trade-offs in terms of the robustness criteria are system-dependent but certain universal features were identified.
Specifically, a set of robust pathways may not combine optimally under nominal condition if the interaction between the control field and the dipole parameter is not strong enough to maximize the transition probability within a specified field duration and amplitude. In the presence of uncertainties, however, these pathways may be associated with a small magnitude of uncertainty such that their expected value and variance are relatively more robust.

On the other hand, the implementation of model-free robust control was demonstrated in a closed feedback loop, where robust optimization over amplitude noise is considered. Two methods of robust control, adaptive GA and traditional GA, were studied in order to reach a solution in a dynamic and static landscape, respectively. Robustness analysis of the solutions obtained reveals that the two algorithms entail different methods of robustness. While a traditional GA converges to a particular solution with the highest expected transition probability, adaptive GA converges to a set of population. The individual solutions within the population obtained in the latter are generally less robust in terms of their expected transition probability than that obtained with traditional GA. They are, however, collectively more robust as the best solution is utilized at different instances of the landscape which changes due to noise. It is important to note that in addition to gaining an understanding of the magnitude of variation in $J$ due to uncertainties in the system and field parameters, the method of asymptotic robustness analysis is also used to determine the number of samples required to achieve an accurate estimation of sampled mean. This is useful in both model-based and model-free robust control when the calculation of expected transition probability using the asymptotic method is computationally expensive. To implement the asymptotic robustness analysis method for large quantum systems, parallel processing can be applied.

We note that several of the Pareto fronts reported in this paper could not have been effectively produced with standard leading order approximations, since inaccuracies can reduce performance measure fidelity if the leading order Taylor approximation for moments 
is used at each step of $\varepsilon(t)$ optimization. For example, the inaccuracy of the Taylor approximation for $\mathrm{E}[J]$ is exaggerated for the most common scenario of Gaussian uncertainty (noise), because the first-order correction is always 0, leaving only the second-order correction to account for the effects of uncertainty.
In addition, as noted above, the standard first-order approximation for var $J$ \cite{nagy2004} is also 0 when $J_{\text{nom}}$ maximized; hence any front containing var $J$ and $J_{\text{nom}}$ cannot be constructed with the standard first-order approximation to the variance. In practice leading order approximations beyond first and second-order approximations \cite{nagy2004, hock2014} are almost never used, due to the need for analytical expressions for each higher order derivative of $J$ and the lack of available numerical methods for their evaluation beyond term-by-term calculation.

The present work has provided a  foundation for the application of evolutionary algorithms to both a) model-based quantum robust control in the presence of Hamiltonian uncertainty and b) model-free quantum robust control in the presence of field noise. In future work, quantum adaptive feedback controllers can be designed which improve the convergence of evolutionary quantum control algorithms by applying model-based robust control to provide an initial guess for model-free learning control and  then iteratively updating Hamiltonian parameter estimates based on the data obtained from learning control. Applications to quantum information processing \cite{hock2014} may also be explored.

Recently, the application of sampling-based learning control to experimental quantum robust control design in the presence of laser field noise has been reported \cite{rabi2018}. This demonstrates the feasibility of applying robust genetic or evolutionary learning control algorithms in the laboratory to achieve robust quantum control, and also motivates  the integration of model-based quantum robust control in the presence of Hamiltonian uncertainty with rrobust experimental learning control. In addition, the methodologies applied herein can be employed to reveal the control mechanisms whereby robust quantum control is achieved in such applications.

\end{document}